\documentclass[pre,twocolumn,floatfix, superscriptaddress,a4paper,nofootinbib]{revtex4}
\usepackage{amsmath,bm,graphicx}
\usepackage{amssymb}
\usepackage{amsfonts}
\usepackage{graphicx}
\usepackage{epstopdf}
\usepackage{amssymb}
\usepackage{mathrsfs}
\usepackage{amsmath}
\usepackage{color}
\usepackage{latexsym}
\usepackage{amsfonts}
\usepackage{hyperref}
\usepackage{subfigure}
\usepackage{wasysym}
\usepackage{dcolumn}
\usepackage{bm}
\usepackage{natbib}
\usepackage{ulem}

\begin{document}

\title{Electroneutrality breakdown for electrolytes embedded in varying-section nanopores}

\author{Paolo Malgaretti}
\email[Corresponding Author: ]{p.malgaretti@fz-juelich.de }
\affiliation{Helmholtz Institute Erlangen-N\"urnberg for Renewable Energy (IEK-11), Forschungszentrum J\"ulich, Erlangen, Germany}

\author{Ignacio Pagonabarraga}
\affiliation{ 
Departament de F\'{\i}sica de la Mat\`eria Condensada, Universitat de Barcelona, Mart\'{\i} i Franqu\'es 1, 08028 Barcelona, Spain  
} 
\affiliation{   
Universitat de Barcelona Institute of Complex Systems (UBICS), Universitat de Barcelona, 08028 Barcelona, Spain 
} 
\author{Jens Harting}
\affiliation{Helmholtz Institute Erlangen-N\"urnberg for Renewable Energy (IEK-11), Forschungszentrum J\"ulich, Erlangen, Germany}
\affiliation{Department of Chemical and Biological Engineering and Department of Physics, Friedrich-Alexander-Universität Erlangen-Nürnberg, Erlangen, Germany}

\date{\today}

\begin{abstract}
We determine the local charge dynamics of a $z-z$ electrolyte embedded in a varying-section channel. 
By means of an expansion based on the length scale separation between the axial and transverse direction of the channel, we derive closed formulas for the local excess charge for both, dielectric and conducting walls, in $2D$ (planar geometry) as well as in $3D$ (cylindrical geometry). 
Our results show that, even at equilibrium, the local charge electroneutrality is broken whenever the section of the channel is not homogeneous for both dielectric and conducting walls as well as for $2D$ and $3D$ channels.
Interestingly, even within our expansion, the local excess charge in the fluid can be comparable to the net charge on the walls. 
We critically discuss the onset of such local electroneutrality breakdown in particular with respect to the correction that it induces on the effective free energy profile experienced by tracer ions. 
\end{abstract}

\maketitle

\section{Introduction}

Understanding the dynamics of electrolytes embedded in varying section pores is crucial for many biological~\cite{Albers} as well as technological applications~\cite{Lyderic_Charlaix,laucirica_nanofluidic_2021}.
For example, ion-channels~\cite{review_Giacomello2022}, plant circulation~\cite{Strook2008}, as well as lymphatic~\cite{Nipper2011} and interstitial~\cite{Wiig2012} transport rely on the active transport of electrolytes across tortuous conduits. 
Moreover, resistive-pulse sensing techniques measure tracer properties during their transport across charged nanopores~\cite{Saleh2003,Ito2004,Heins2005,Arjmandi2012} and energy-harvesting devices exploit the physical properties of confined electrolytes~\cite{Brogioli2009,VanRoij2014,VanRoij2017}.

From a theoretical perspective, the usual approach is based on the Poisson-Boltzmann theory, within which the ions are regarded as point-like particles whose distribution fulfills the Boltzmann weight that is eventually determined by the solution of the Poisson equation. Such an approach can recover the dynamics of electrolytes confined between charged plates provided that the ions are sufficiently diluted and the surface charges are not too large. Within such conditions, the Poisson-Boltzmann theory recovers the global electroneutrality, namely the total charge in the liquid phase matches the one on the plates. 
\begin{figure}[t]
 \includegraphics[scale=0.41]{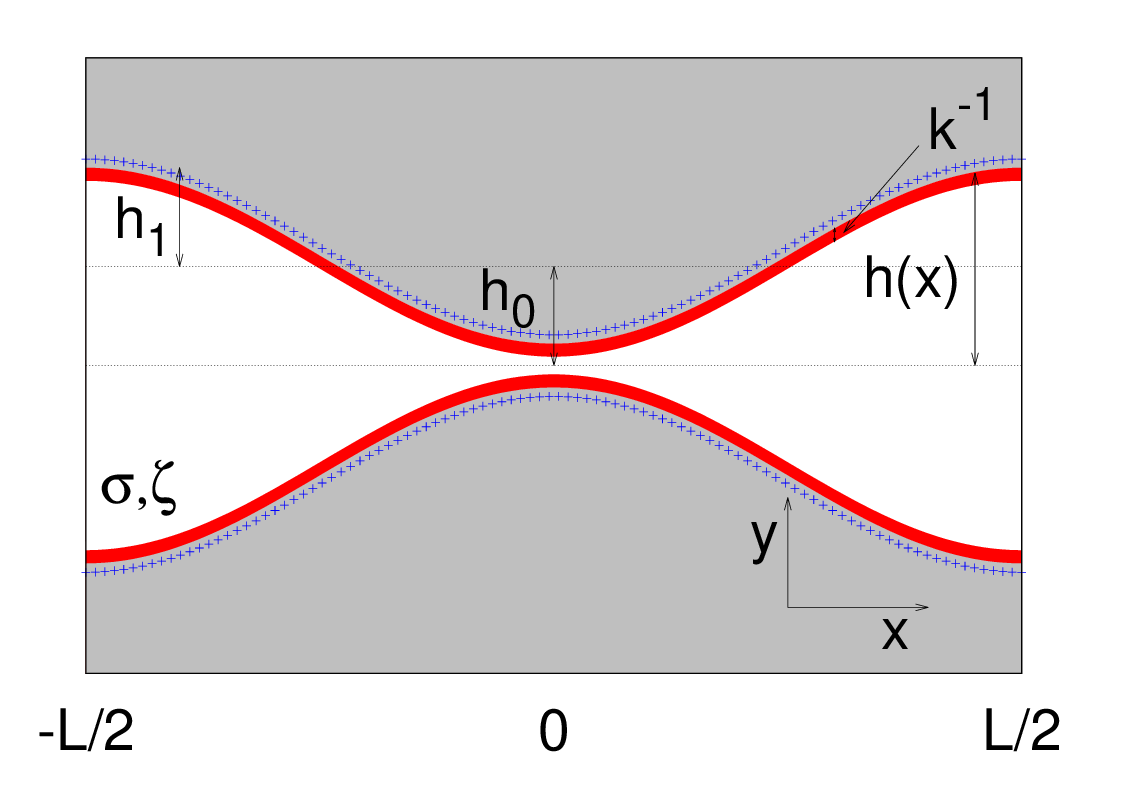}
 \caption{Schematic view of the system: channel walls are in grey, the local surface charge is represented by blue crosses whereas the width of the Debye double layer is in red.
 }
 \label{fig:scheme}
\end{figure}
However, recent experimental~\cite{luo_electroneutrality_2015} and theoretical~\cite{colla_charge_2016,keshavarzi_effect_2020,Bazant2020,Bazant2021} results show that when the plates are approaching at distances $\lesssim 10$nm the reduced space and the steric and electrostatic interactions between the dissolved ions may lead to a breakdown of the global electroneutrality. This means that the net charge in the liquid between the plates does not balance the charge on the plates. However, since the electrolyte is typically in contact with a reservoir, the eventual global electroneutrality is attained when accounting for the charge distribution outside the channel~\cite{colla_charge_2016}.
While global electroneutrality breakdown is associated with a significant increase in the total free energy of the system~\cite{lozada-cassou_violation_1996}, local rearrangements of the charge may occur such that the electroneutrality is fulfilled globally but not locally, as it has been recently predicted for macroion solutions~\cite{gonzalez-calderon_violation_2021}. 

In particular, when the section of the confining vessel is not constant, novel dynamical regimes appear. Indeed, asymmetric pores have been used to rectify ionic currents~\cite{Siwy2005,hanggi,Wegrowe2016}, 
as well as to realize highly sensitive dopamine-responsive iontronic devices~\cite{Azzaroni2020}. Moreover, recirculation and local electroneutrality breakdown have been reported for electrolytes confined between corrugated walls~\cite{Malgaretti2014,Chinappi2018} and the variation in channel section can tune their permeability~\cite{Malgaretti2015,Malgaretti2016} and even enhance the effective transport coefficients~\cite{Malgaretti2019}.

Accordingly, the question arises about the interplay between the geometry of the pore and the electroneutrality breakdown. In order to be able to efficiently explore the parameter space we aim at an analytical approach that can provide closed formulas capturing the dependence of  electroneutrality breakdown on the parameters characterizing the system. Accordingly, we exploit the (linearized) Poisson-Boltzmann theory. 
Interestingly, our results show that even in the simplest scenario, i.e., at equilibrium and within the Debye-H\"uckel approximation, the interplay between the (varying) local section of the channel and the electrostatic forces leads to a breakdown of local electroneutrality and to the onset of an inhomogeneous excess charge for both dielectric and conducting channel walls in planar ($2D$) and cylindrical ($3D$) geometries.  
Interestingly, upon tuning the parameters, the local excess charge can be as large as the local charge on the walls. Once integrated along the transverse direction the local excess charge can be expressed via a multipole expansion whose leading term is a quadrupole.  
Finally, such an insight can be used to predict the corrections to the effective free energy profile experienced by a tracer ion induced by the local excess charge. 


\section{$2D$ Model}
In the following, we analyze the case of a channel whose half-section $h(x)$ (see Fig.\ref{fig:scheme})
\begin{align}
    h(x)=h_0\left(1-h_1\cos(2\pi x/L)\right)
\end{align}
varies solely along the $x$ direction. For later use, we introduce the entropic barrier
\begin{equation}
  \Delta S=\ln\left[\frac{h_{\text{max}}}{h_\text{min}}\right].
 \end{equation}
Here, $h_{\text{max}}\equiv h_0(1+h_1)$ and $h_\text{min}\equiv h_0(1-h_1)$ are, respectively, the maximum and minimum channel sections.
and which is filled with a $z-z$ electrolyte. The electrostatic potential $\phi$ inside the channel is determined by solving the (linearized) Poisson-Boltzmann  equation,
\begin{equation}
\nabla_{x,y}^{2}\phi(\mathbf{x})=k^2 \phi(\mathbf{x}),
\label{eq:poisson}
\end{equation}
where $k^{-1}=\lambda$ is the inverse Debye length. 

In order to get an analytical insight into the solution of Eq.~(\ref{eq:poisson}) we exploit the lubrication approximation, based on a  separation between longitudinal and transverse length scales, for which we assume that the changes of $\phi$ along the longitudinal direction are much smaller than those along the transverse direction. In the system under study, the longitudinal length scale is captured by $L$ which is the period of the channel section, whereas the transverse length scales are the channel average section $h_0$ and the Debye length $\lambda$.
Hence, we identify $\lambda/L$ as the small parameter. 
Accordingly, we rewrite Eq.~\eqref{eq:poisson} as
\begin{equation}
\frac{\lambda^2}{L^2} \frac{\partial^2 \phi}{\partial^2 x_*}+\frac{\partial^2 \phi}{\partial y_*^{2}}=k^2 \phi.
\label{eq:poisson2}
\end{equation}
In the last expression, we have highlighted that the first term on the left-hand side is of $\mathcal{O}\left(\frac{\lambda}{L}\right)^2$ as compared to the second one. Accordingly, in the lubrication approximation, we treat the dependence on the longitudinal coordinate parametrically.

The solution of Eq.~(\ref{eq:poisson}) is governed by the boundary conditions on the channel walls. For conducting channel walls the potential equals the zeta potential $\zeta$: 
\begin{equation}
\phi_0(x,h(x))=\zeta\, 
\end{equation}
Alternatively, for insulating channel walls, we have that 
\begin{eqnarray}
-\nabla \phi\cdot\mathbf{n}|_{y=\pm h(x)}&=&\mathbf{E}\cdot\mathbf{n}=\pm \frac{\sigma}{\epsilon}, \label{eq:BC-n}
\end{eqnarray}
where $\mathbf{n}$ is the unit vector normal at the channel walls and pointing inside the channel.

In particular, calling $\alpha=\arctan(\partial_xh(x))$ the local slope of the channel walls and focusing on the upper channel wall (the same calculations can be re-derived for the lower one) we have:
\begin{eqnarray}
 \mathbf{n}&=&\left(\begin{array}{c}
             \sin\alpha\\
             -\cos\alpha
            \end{array}\right)
\end{eqnarray}
Accordingly, we get 
\begin{align}
 \left[-\partial_x\phi(x,y)\sin\alpha+\partial_y\phi(x,y)\cos\alpha\right]_{y=\pm h(x)}&=\pm \frac{\sigma}{\epsilon}.\label{eq:BC-n-1}
\end{align}
In order to determine the boundary condition at different orders in the lubrication expansion we recall that 
\begin{align}
\alpha \simeq \partial_x h(x) +\mathcal{O}\left(\frac{h_0}{L}\right)^3
\end{align}
and hence we have that at leading order in lubrication the boundary condition reads
\begin{align}
\partial_y \phi_0(x,y)|_{y=\pm h(x)}=\pm \frac{\sigma}{\epsilon},
\label{eq:BC-1}
\end{align}
whereas at the second order (i.e. next leading order) we have 
\begin{align}
\partial_y &\phi_2(x,y)|_{y=\pm h(x)}=\\ 
&\!\!\!=\left[\frac{1}{2}\partial_y \phi_0(x,y)(\partial_x h(x))^2+\partial_x\phi_0(x,y)\partial_x h(x)\right]_{y=\pm h(x)}.\nonumber
\label{eq:BC-2}
\end{align}
At equilibrium, we use the following \textit{ansatz} for the linearized ionic number densities:
\begin{subequations}\label{eq:def-dens-lin}
\begin{eqnarray}
\tilde{\rho}_{+}(x,y) & = & \rho_{+}(x)\left(1-\beta ez\phi(x,y)\right)\\
\tilde{\rho}_{-}(x,y) & = & \rho_{-}(x)\left(1+\beta ez\phi(x,y)\right)
\end{eqnarray}
\end{subequations}

\section{$2D$ First order lubrication approximation}
At leading order in the lubrication expansion, namely at order $\mathcal{O}(h_0/L)^0$, the ionic number densities read
\begin{subequations}\label{eq:def-dens-lin-0}
\begin{eqnarray}
\tilde{\rho}_{+,0}(x,y) & = & \rho_{+,0}(x)\left(1-\beta ez\phi_0(x,y)\right),\\
\tilde{\rho}_{-,0}(x,y) & = & \rho_{-,0}(x)\left(1+\beta ez\phi_0(x,y)\right)
\end{eqnarray}
\end{subequations}
and the Poisson equation reads
\begin{equation}
\partial_{y}^{2}\phi_0(x,y)= -\frac{ze}{\epsilon}\left[\psi_0(x)-\varphi_0(x)\beta ze\phi_0(x,y)\right],
\label{eq:poisson1-0}
\end{equation}
where we have introduced the notation
\begin{subequations}\label{eq:def-phi-psi0}
\begin{align}
 \varphi_0(x)&=\rho_{+,0}(x)+\rho_{-,0}(x),\\
 \psi_0(x)&=\rho_{+,0}(x)-\rho_{-,0}(x).
\end{align}
\end{subequations}
The solution of Eq.~(\ref{eq:poisson1-0}) is then given by
\begin{equation}
\phi_0(x,y)=A_0(x)\cosh\left(k_0(x)y\right)+\frac{ze}{\epsilon k_0^{2}(x)}\psi_0(x),
\label{eq:phi-0}
\end{equation}
where 
\begin{equation}
 k_0(x)=\sqrt{\frac{\beta (ze)^2}{\epsilon}\varphi_0(x)}
 \label{eq:def-k}
\end{equation}
is the, zeroth-order, inverse Debye length and $A_0$ is determined by the boundary conditions imposed by the channel walls. For conducting channel walls the potential equals the zeta potential $\zeta$ at the walls, $\phi_0(x,h(x))=\zeta$. Hence, we have 
\begin{align}
\phi_0(x,y)&=\zeta\frac{\cosh\left(k_0(x)y\right)}{\cosh\left(k_0(x)h(x)\right)}+\nonumber\\
&+\frac{ze}{\epsilon k^{2}}\psi_0(x)\left(1-\frac{\cosh\left(k_0(x)y\right)}{\cosh\left(k_0(x)h(x)\right)}\right).
\end{align}
Alternatively, for insulating channel walls and for smoothly varying-channels, $\partial_x h(x)\ll 1$,  the boundary condition can be expressed as~\cite{Malgaretti2015} 
\begin{equation}
 \left.\frac{\partial\phi_0}{\partial y}\right|_{y=\pm h(x)}=\pm\frac{\sigma}{\epsilon}\left(1+\frac{1}{2}\left(\partial_xh(x)\right)^2\right)+\mathcal{O}(\partial_x h(x))^3,
 \label{eq:approx-BC}
\end{equation}
where we have substituted $\alpha\simeq\partial_xh(x)$ in Eq.~(\ref{eq:approx-BC}). Accordingly, using Eq.~(\ref{eq:approx-BC}) the electrostatic potential reads 
\begin{equation}
\phi_0(x,y)=\frac{\sigma}{\epsilon k_0(x)}\frac{\cosh\left(k_0(x)y\right)}{\sinh\left(k_0(x)h(x)\right)}+\frac{ze}{\epsilon k_0^{2}(x)}\psi_0(x)\,.
\label{eq:electric_potential-0}
\end{equation}
At equilibrium, in order to solve for $\rho_{+,0}(x),\rho_{-,0}(x)$, we impose that the electrochemical potential is constant~\cite{RusselBook}:
\begin{equation}
 \mu_\pm(x,y)=k_BT \ln \tilde\rho_\pm(x,y) \pm ze\phi(x,y)=\bar\mu_\pm.
 \label{eq:def-mu}
\end{equation} 
Disregarding terms of order $\mathcal{O}(\phi_0)^2$, by plugging Eqs.~(\ref{eq:def-dens-lin-0}) into Eq.~(\ref{eq:def-mu}) leads to
\begin{align}
 \rho_{\pm,0}(x)=e^{\beta \bar\mu_\pm}
 \label{eq:rho-pm-0}
\end{align}
and the constant values of $\rho_{\pm,0}$ are set by the equilibrium chemical potentials $\bar\mu_\pm$. For $z-z$ electroneutral systems we have $\mu_+=\mu_-$ and therefore 
\begin{equation}
 \rho_{+,0}=\rho_{-,0}\,.
 \label{eq:psi-0}
\end{equation}
This implies
\begin{align}
 \varphi_0(x)&=2\rho_0,\label{eq:varphi0}\\
 \psi_0(x)&=0.\label{eq:psi0}
\end{align}
In particular, Eq.~(\ref{eq:varphi0}) leads to
\begin{equation}
k_0(x)=k_0,
\label{eq:k0}
\end{equation}
i.e., the Debye length is constant along the channel. Finally, the electrostatic potential for conducting channel walls reads
\begin{equation}
 \phi^\zeta_0(x,y)=\zeta\frac{\cosh\left(k_0y\right)}{\cosh\left(k_0h(x)\right)},
 \label{eq:electric_potential-cond}
\end{equation}
whereas for insulating walls it is
\begin{equation}
\phi^\sigma_0(x,y)=\frac{\sigma}{\epsilon k_0}\frac{\cosh\left(k_0y\right)}{\sinh\left(k_0h(x)\right)}.
\label{eq:electric_potential-0-1}
\end{equation}
Substituting Eq.~(\ref{eq:electric_potential-cond}) or Eq.~(\ref{eq:electric_potential-0-1}) into Eq.~(\ref{eq:poisson1-0}) and using Eq.~(\ref{eq:psi0}) the net local charge reads, respectively,
\begin{align}
 &q^\zeta_0(x)=-\!\!\!\!\int\limits_{-h(x)}^{h(x)}\!\!\!\epsilon k_0^2 \zeta\frac{\cosh\left(k_0y\right)}{\cosh\left(k_0h(x)\right)}dy=-2\epsilon k_0 \zeta \tanh(k_0 h(x)),\label{eq:q0-zeta}\\
 &q^\sigma_0(x)=-\!\!\!\!\int\limits_{-h(x)}^{h(x)}\!\!\!\epsilon k_0^2 \frac{\sigma}{\epsilon k_0}\frac{\cosh\left(k_0y\right)}{\sinh\left(k_0h(x)\right)}dy=-2\sigma,
 \label{eq:q0}
\end{align}
which recovers the local electroneutrality of the system\footnote{We note that, at zeroth order in lubrication, the surface charge at each conducting wall can be obtained by $\sigma^\zeta=\pm\epsilon \partial_y\phi_0^\zeta(x,y)|_{y=\pm h(x)}=\epsilon k_0 \zeta \tanh(k_0 h(x))$.}.
Equations~(\ref{eq:k0})-(\ref{eq:q0}) show that at leading order in lubrication, there is no correction to the Debye length, electrostatic potential or local charge and ionic density induced by the geometrical confinement. 
Such a tight relationship between the surface charge and the charge in the liquid phase is carved into the Debye-H\"uckel equation. In fact, multiplying by $-\epsilon$ and integrating the Debye-H\"uckel equation at first order in lubrication along the transverse coordinate leads to 
\begin{align}
-\epsilon \int_{-h(x)}^{h(x)}\partial^2_y \phi(x,y) dy = -\epsilon \int_{-h(x)}^{h(x)}k_0^2 \phi(x,y) dy.
\label{eq:DB}
\end{align}
After integrating it follows that
\begin{align}
-\epsilon\partial_y   \phi(x,y)|^{h(x)}_{-h(x)}=2\sigma=q(x),
\end{align}
where we have identified that the right-hand side of Eq.~\eqref{eq:DB} is indeed the total charge in the fluid and the left-hand side corresponds to minus the surface charge. 
Therefore, in order to check if local electroneutrality can be broken also at equilibrium, we need to account for higher order corrections in the lubrication approximation.  

\section{$2D$ Higher order corrections}
It is clear that, at order $\mathcal{O}(\frac{h_0}{L})$, Eq.~(\ref{eq:poisson2}) jointly with the boundary conditions, Eq.~(\ref{eq:approx-BC}) leads to $\phi_1(x,y)=0$.
Therefore, the next leading order is $\mathcal{O}(\frac{h_0}{L})^2$. 
Accordingly, the ansatz of the ionic number densities reads
\begin{subequations}\label{eq:def-rho2}
\begin{align}
\tilde{\rho}_{+,2}(x,y) & = \rho_{+,2}(x)\left(1-\beta ez\phi_0(x,y)\right)-\rho_{+,0}\beta ez\phi_2(x,y),\label{eq:def-rho+2}\\
\tilde{\rho}_{-,2}(x,y) & = \rho_{-,2}(x)\left(1+\beta ez\phi_0(x,y)\right)+\rho_{-,0}\beta ez\phi_2(x,y).\label{eq:def-rho-2}
\end{align}
\end{subequations}
At equilibrium the chemical potential is homogeneous. Therefore, since the zeroth order contribution to the chemical potential is already fulfilling the equilibrium conditions, see Eq~(\ref{eq:def-mu}), we have that 
\begin{equation}
 \mu_{\pm,2}(x,y)=0.
 \label{eq:equil-cond-2}
\end{equation}
The expression for the second order contribution to the local chemical potential reads
\begin{equation}
 \mu_{\pm,2}(x,y)=k_B T\frac{\tilde{\rho}_{\pm,2}(x,y)}{\rho_{\pm,0}}\pm ze \phi_2(x,y).
\end{equation}
Using Eqs.~(\ref{eq:rho-pm-0}),(\ref{eq:psi-0}),(\ref{eq:def-rho+2}),(\ref{eq:def-rho-2}) and keeping only terms linear in the electrostatic potential, Eqs.~(\ref{eq:equil-cond-2}) leads to
\begin{subequations}\label{eq:equil-2}
\begin{align}
 \rho_{+,2}(x)\left(1-\beta ze\phi_0(x,y)\right)&=0,\\
 \rho_{-,2}(x)\left(1+\beta ze\phi_0(x,y)\right)&=0,
\end{align}
\end{subequations}
from which we obtain
\begin{subequations}\label{eq:rho2}
\begin{align}
\rho_{+,2}(x)&=0,\\
\rho_{-,2}(x)&=0.
\end{align}
\end{subequations}
Accordingly, the Poisson equation reads
\begin{equation}
\partial_{x}^{2}\phi_0(x,y)+\partial_{y}^{2}\phi_2(x,y)= k_0^2\phi_2(x,y)\,.
\label{eq:DH-2}
\end{equation}
We remark that, due to Eqs.~\eqref{eq:rho2}, there is no second order correction to the Debye length, which indeed is clear from Eq.~\eqref{eq:DH-2}. The solution of Eq.~\eqref{eq:DH-2}, using Eq.~(\ref{eq:phi-0}), reads
\begin{align}
 \phi&_2(x,y)=A_2(x)\cosh(k_0 y)+\frac{1}{4k_0^2}\partial^2_x A_0(x)\xi(x,y),
\label{eq:sol-phi-2}
\end{align}
with
\begin{equation}
 \xi(x,y)=\cosh(k_0 y)-2k_0 y\sinh(k_0 y),
\end{equation}
where $A_0(x)$ is the zeroth--order integration constant and it is determined by the following boundary conditions:
\begin{equation}
 A^\zeta_0(x)=\frac{\zeta}{\cosh(k_0 h(x))}
\end{equation}
for conducting walls and 
\begin{equation}
 A^\sigma_0(x)=\frac{\sigma}{\epsilon k_0}\frac{1}{\sinh(k_0 h(x))}
\end{equation}
for dielectric walls.
Finally, $\phi_{2}(x,y)$ is determined by imposing the boundary conditions at the channel walls.
For conducting channel walls, at order $\mathcal{O}\left(\frac{h_0}{L}\right)^2$, the boundary condition reads
\begin{equation}
 \phi_2(x,\pm h(x))=0.
\end{equation}
This leads to 
\begin{equation}
 \!\phi^\zeta_2(x,y)=\!\frac{\partial^2_x A^\zeta_0(x)}{4k_0^2}\left[\xi(x,y)-\frac{\cosh(k_0 y)}{\cosh(k_0h(x))}\xi(x,h(x))\right].
 \label{eq:phi2_zeta}
\end{equation}
In contrast, for dielectric channel walls the boundary conditions, at order $\mathcal{O}(\frac{h_0}{L})^2$ is given by Eq.~\eqref{eq:approx-BC}
which leads to (similar results can be obtained for $y=-h(x)$)
\begin{align}
 \partial_y\phi_2|_{y=h(x)}=&\frac{1}{2}\frac{\sigma}{\epsilon}\left(\partial_xh(x)\right)^2\nonumber\\
 &+\cosh(k_0 h(x))\partial_x A_0(x)\partial_x h(x),
 \label{eq:approx-BC-2}
\end{align}
from which we obtain
\begin{align}
A^\sigma_2(x) &= \frac{\sigma}{2 k \epsilon}\frac{\left(\partial_xh(x)\right)^2}{\sinh(k_0 h(x))}\nonumber\\
&+\frac{\cosh(k_0 h(x))}{k_0\sinh(k_0 h(x))}\partial_x A^\sigma_0(x)\partial_x h(x)\\
& +\frac{\partial^2_x A^\sigma_0(x)}{4k_0^2}\left[1+2k_0 h(x)\frac{\cosh(k_0 h(x))}{\sinh(k_0 h(x))}\right]\nonumber
\end{align}
and accordingly we get
\begin{align}
 &\phi^\sigma_2(x,y)=\frac{1}{2}\phi^\sigma_0(x,y)\left(\partial_x h(x)\right)^2\nonumber\\
 &+\frac{\epsilon}{\sigma}\phi^\sigma_0(x,y)\cosh(k_0h(x))\partial_x A^\sigma_0(x)\partial_x h(x) \nonumber\\
 &+\frac{\partial^2_x A^\sigma_0(x)}{2k_0^2}\left[\left(k_0h(x)\frac{\cosh(k_0 h(x))}{\sinh(k_0h(x))}+1\right)\cosh(k_0 y)\right.\nonumber\\
 &-k_0y \sinh(k_0 y)\Big].
 \label{eq:phi2_sigma}
\end{align}
\section{$2D$ local electroneutrality breakdown}

\begin{figure}[t]
 \includegraphics[scale=0.5]{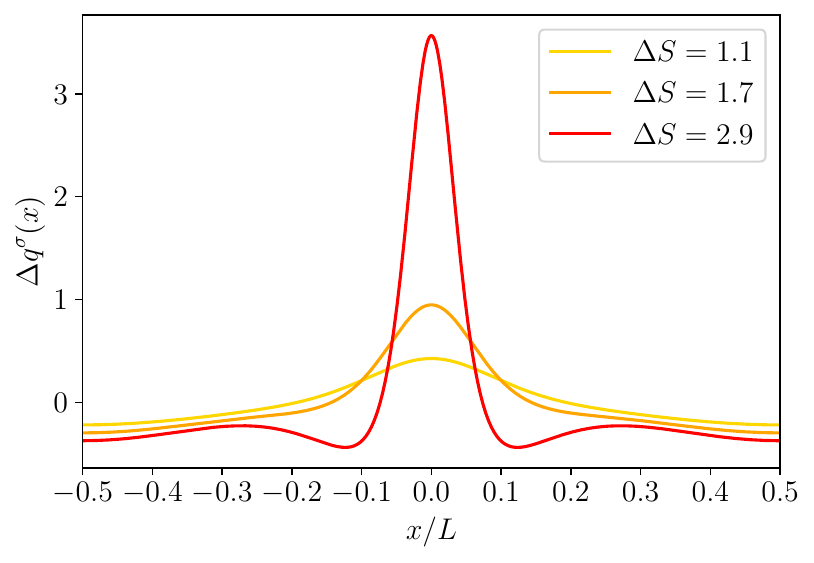}
 \caption{$2D$ Dielectric channel. Local excess charge $\Delta q(x)$ as a function of the normalized position $x/L$ along the channel axis for different values of $\Delta S$ as detailed in the legend with $h_0/L=0.1$ and $kh_0=1$. }
 \label{fig:q_Eq}
\end{figure}

\paragraph{Dielectric channel walls}
Using Eq.~(\ref{eq:phi2_sigma}) we can calculate the corrections to the local charge for dielectric channel walls\footnote{We recall that $\int_{-h(x)}^{h(x)} k^2y \sinh(ky)dy =  2 h k \cosh(h k) - 2 \sinh(h k)$.}:
\begin{align}
 q_2^\sigma(x) &=-\epsilon k_0^2 \int_{-h(x)}^{h(x)}\phi^\sigma_2(x,y)dy\nonumber\\
 &=  -\sigma\left(\partial_x h(x)\right)^2 - 2\epsilon \cosh(k_0h(x))\partial_x A^\sigma_0(x)\partial_x h(x)  \nonumber\\
 &-2\epsilon\frac{\partial^2_x A^\sigma_0(x)}{k_0}\sinh(k_0 h(x)) 
 \label{eq:q2-sig}
\end{align}
For a dielectric channel, at second order in lubrication $\mathcal{O}(h_0/L)^2$, the charge per unit area on the channel walls reads
\begin{equation}
 q^\sigma_w(x)=\sigma\left(1+\frac{1}{2}\left(\partial_x h(x)\right)^2\right).
 \label{eq:sig-wall-2}
\end{equation}
Accordingly, we can define the excess charge
\begin{align}
 \Delta q^\sigma(x)&=\frac{q^\sigma_0+q^\sigma_2(x)+2q^\sigma_w(x)}{2 |q_{0}^\sigma|}. \label{eq:Delta-q}
\end{align}
Using Eqs.~\eqref{eq:q0},\eqref{eq:q2-sig},\eqref{eq:sig-wall-2}, at second order in lubrication, the last expression reduces to 
\begin{align}
 \Delta q^\sigma(x)\simeq & -\frac{\epsilon}{k_0 \sigma}\left[\partial_x\sinh(k_0h(x)) \partial_x A^\sigma_0(x)\right.\nonumber\\
 &\left. +\partial^2_x A^\sigma_0(x)\sinh(k_0 h(x))\right]\,.
\end{align}
This can be rewritten as 
\begin{align}
 \Delta q^\sigma(x)\simeq - \frac{\epsilon}{k_0 \sigma}\partial_x\left[\sinh(k_0 h(x))\partial_x A_0^\sigma (x)\right]\,.
 \label{eq:Deltq-q}
\end{align}
Fig.~\ref{fig:q_Eq} shows the dependence of $\Delta q^\sigma$ on the longitudinal position. Interestingly, there is an excess charge with respect to the case of planar channel walls close to the channel bottleneck at $x/L=0.5$ and a charge depletion in the remainder of the channel. Interestingly, the distribution of the excess charge is symmetric about the center of the channel and, in a far--field expansion, it leads to the onset of a net quadrupolar contribution.

\begin{figure*}[t]
 \includegraphics[scale=0.41]{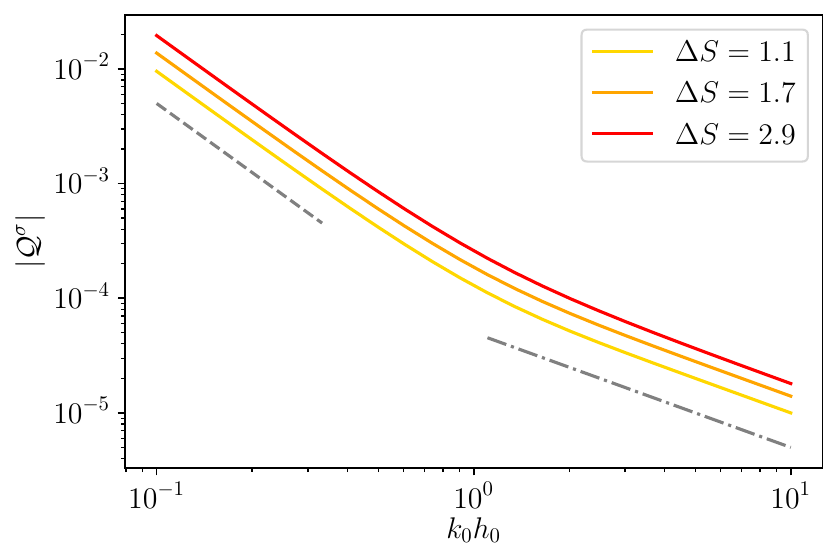}
 \includegraphics[scale=0.41]{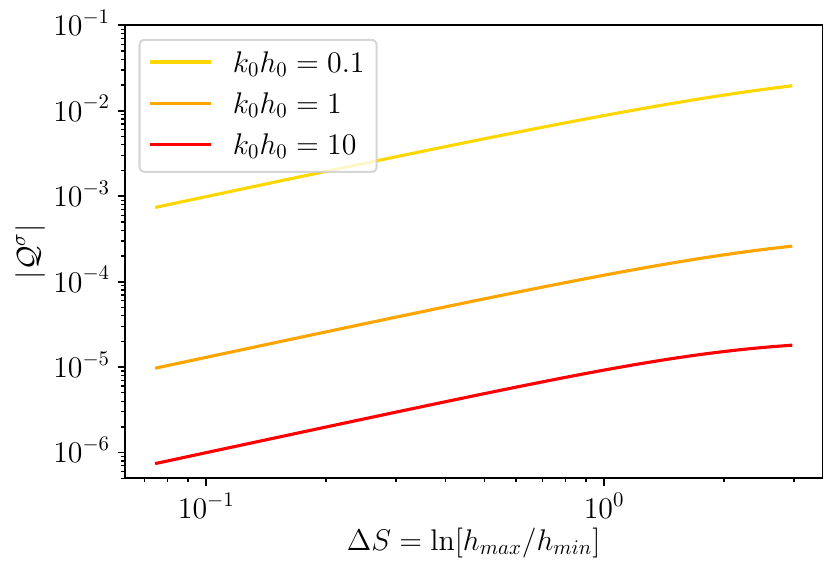}
 \includegraphics[scale=0.41]{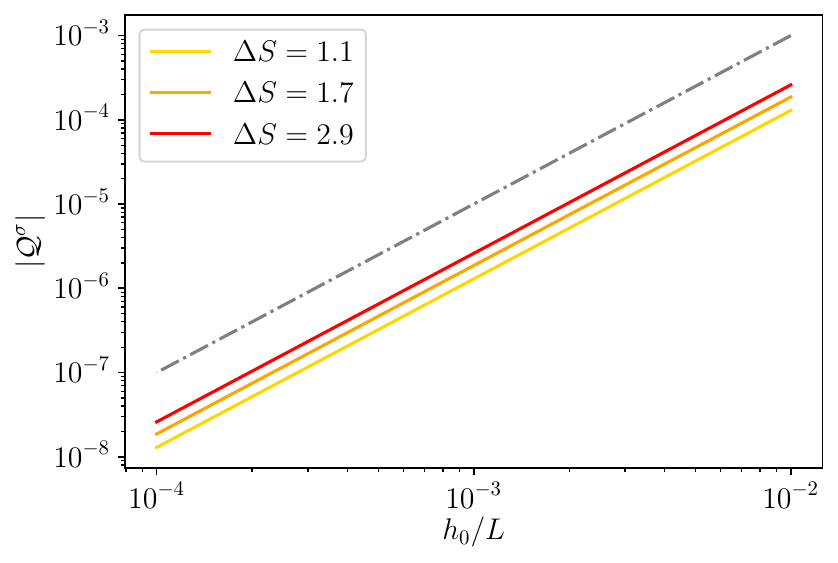}
  \includegraphics[scale=0.41]{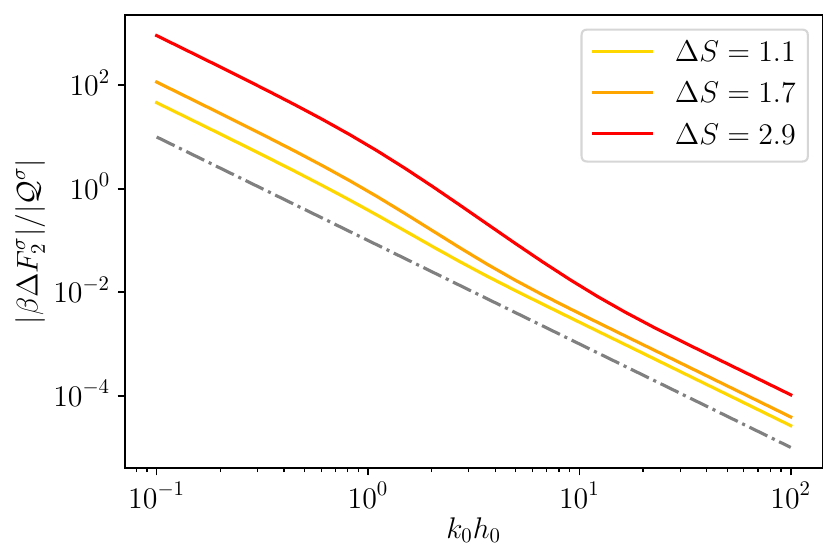}
 \includegraphics[scale=0.41]{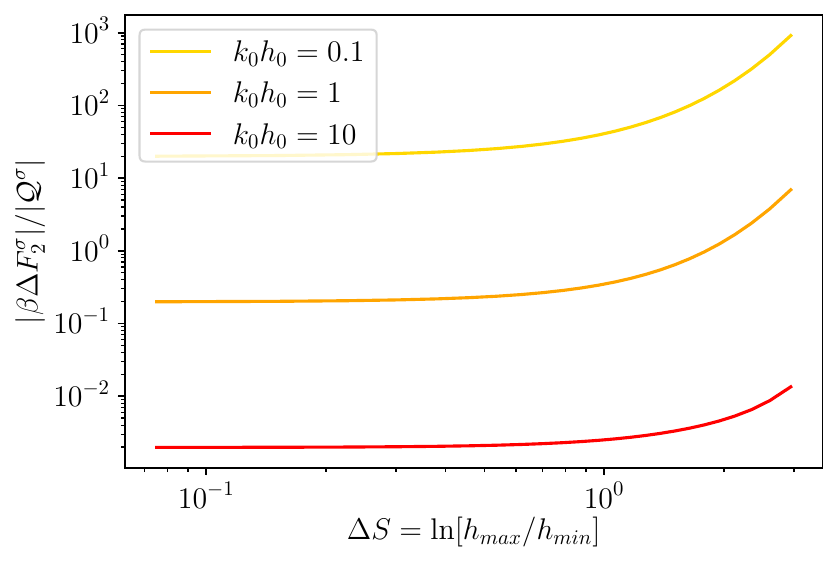}
 \includegraphics[scale=0.41]{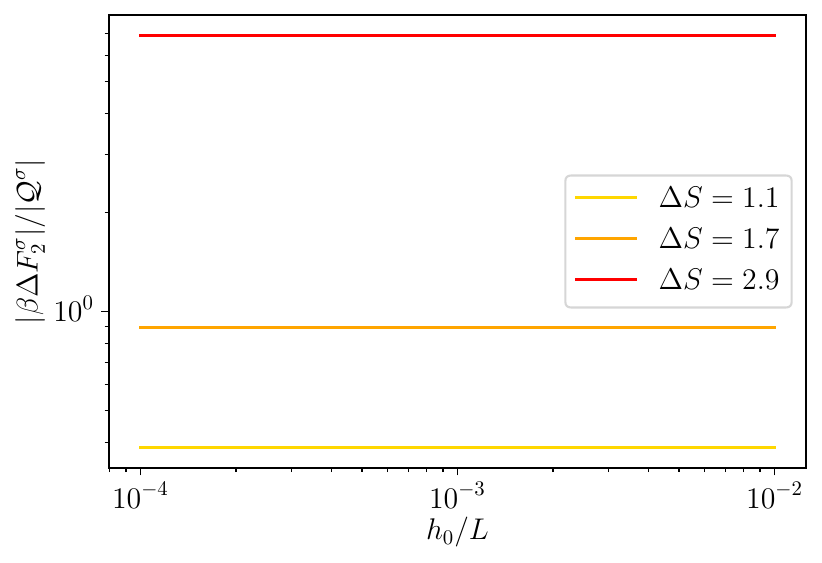}
 \caption{$2D$ Dielectric channel. Top left: absolute value of the quadrupole moment, as defined in Eq.~\eqref{eq:QQ}, $\mathcal{Q}^\sigma$ as a function of $k_0 h_0$ for different values of $\Delta S$ as reported in the legend and with $h_0/L=0.01$. The grey dashed line is proportional to $\propto (k_0 h_0)^{-1}$ and the dashed-dotted one to $\propto (k_0 h_0)^{-2}$. 
 Top center: absolute value of the quadrupole moment, $\mathcal{Q}^\sigma$, as a function of $\Delta S$ for different magnitudes of $kh_0$ as reported in the legend and with $h_0/L=0.1$. The thin grey dotted line is a guide for the eye and it is proportional to $\Delta S^2$. 
 Top right: absolute value of the quadrupole momentum, $\mathcal{Q}^\sigma$, as a function of $h_0/L$ for different magnitudes of $\Delta S$ as reported in the legend and with $kh_0 = 1$. The thin grey dotted line is a guide for the eye and it is proportional to $(h_0/L)^2$. 
 Bottom: ratio $\Delta F^\sigma_2/\mathcal{Q}^\sigma$ for the same values of the parameters of the corresponding top panel.}
 \label{fig:q_Eq-2}
\end{figure*}
We remark that when the local excess charge $\Delta q^\sigma(x)$ is integrated over the channel period it leads to a vanishing contribution, i.e., global electroneutrality is retrieved. 
We quantify the magnitude of the local electroneutrality breakdown by computing the dimensionless quadrupolar moment
\begin{align}
\mathcal{Q}^\sigma = \frac{1}{L}\int_{-\frac{L}{2}}^{\frac{L}{2}}\frac{x^2}{L^2}\Delta q^\sigma(x) dx.
\label{eq:QQ}
\end{align}  
The top panel of Fig.~\ref{fig:q_Eq-2} shows the dependence of $\mathcal{Q}^\sigma$ on $k_0 h_0$. Interestingly, $\mathcal{Q}^\sigma$ shows a twofold scaling: for smaller values of $k_0 h_0$, $\mathcal{Q}^\sigma$ decays as $k_0 h_0^{-2}$, whereas for larger values of $k_0 h_0$ it decays as $k_0 h_0^{-1}$. In particular, the crossover between the two regimes is around $k_0 h_0\simeq 1$, i.e., when the Debye length is comparable to the average channel section. 
The central panel of Fig.~\ref{fig:q_Eq-2} shows that $\mathcal{Q}^\sigma$ grows almost linearly with the entropic barrier $\Delta S$. Finally, the bottom panel of Fig.~\ref{fig:q_Eq-2} shows that $\mathcal{Q}^\sigma$ depends quadratically on $h_0/L$. This is expected since these results were derived at the second order in lubrication. 

\begin{figure}[t]
 \includegraphics[scale=0.5]{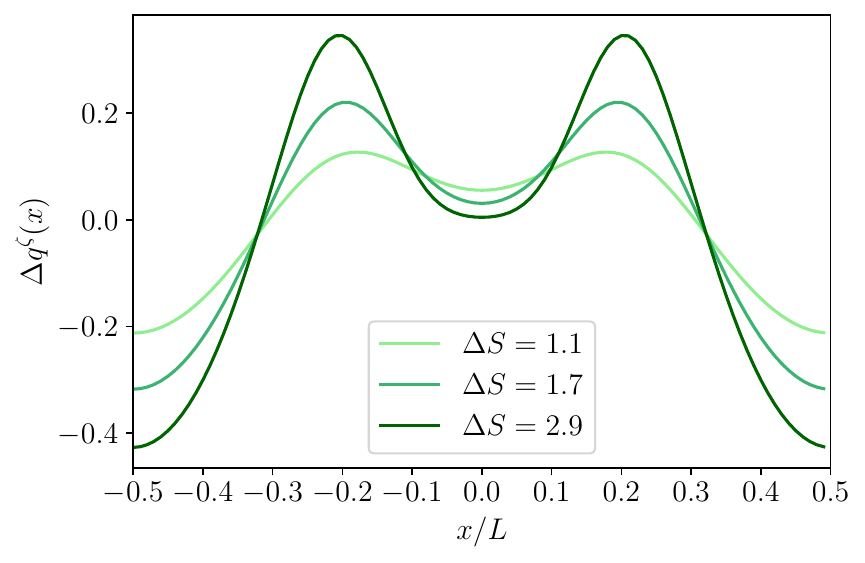}
 \caption{$2D$ Conducting channel. Local excess charge $\Delta q(x)$  as a function of the normalized position $x/L$ along the channel axis for different values of $\Delta S$ as deailed in the legend with $h_0/L=0.1$ and $kh_0=1$. }
 \label{fig:q_Eq-z}
\end{figure}

\paragraph{Conducting channel walls}
Using Eq.~(\ref{eq:phi2_zeta}) we can calculate the corrections to the local charge for conducting channel walls\footnote{We recall that $\int_{-h(x)}^{h(x)} k^2y \sinh(ky)dy =  2 h k \cosh(h k) - 2 \sinh(h k)$}:
\begin{align}
 q_2^\zeta(x) &=-\epsilon k_0^2 \int_{-h(x)}^{h(x)}\phi^\zeta_2(x,y)dy\nonumber\\ 
 &=\epsilon\frac{\partial^2_x A^\zeta_0(x)}{k_0}\left[\frac{k_0 h(x)}{\cosh(k_0 h(x)}-\sinh(k_0 h(x))\right]
  \label{eq:q2-zeta}
\end{align}
For a conducting channel, we first define the effective surface charge as 
\begin{align}
q_w^\zeta(x) \equiv -\epsilon \nabla \phi^\zeta(x,y)|_{y=h(x)}\cdot \mathbf{n},
\end{align}
which up to second order contributions in lubrication reads as
\begin{align}\label{eq:qw_z-2}
 q_w^\zeta(x) &\simeq -\epsilon \left[\partial_x \phi_0^\zeta (x,y)\partial_x h(x)\right.\\
&\left.-\partial_y\phi_0(x,y)\left(1-\frac{1}{2}(\partial_xh(x))^2\right) -\partial_y\phi_2(x,y)\right]_{y=h(x)}. \nonumber
\end{align}
This can be rewritten as 
\begin{align}
q_w^\zeta(x)  \simeq & -\epsilon \left[\cosh(k_0 h(x))\partial_x A_0^\zeta (x) \partial_x h(x)\right.\nonumber\\
& -\zeta k_0 \text{tanh}(k_0 h(x))\left(1-\frac{1}{2}(\partial_xh(x))^2\right)\label{eq:app_q_W-zeta}\\
&\left.+\frac{\partial_x^2 A_0^\zeta(x)}{2 k_0}\left(\frac{k_0 h(x)}{\cosh(k_0 h(x)}+\sinh(k_0 h(x))\right) \right].\nonumber
\end{align}
We recall that the last expression accounts for the surface charge density along the surface of the channel. However, when comparing the surface charge to that in the liquid phase we have to account for the fact that the latter is per unit length $dx$ along the longitudinal axis of the channel. Accordingly, when computing the local electroneutrality we have to multiply Eq.~\eqref{eq:app_q_W-zeta} by the local area $\simeq 1+\frac{1}{2}(\partial_x h(x))^2$. We then get
\begin{align}
q_w^\zeta(x) & \simeq \epsilon \left[\zeta k_0 \text{tanh}(k_0 h(x))-\cosh(k_0 h(x))\partial_x A_0^\zeta (x) \partial_x h(x)\right.\nonumber\\
&\left. -\frac{\partial_x^2 A_0^\zeta(x)}{2 k_0}\left(\frac{k_0 h(x)}{\cosh(k_0 h(x)}+\sinh(k_0 h(x))\right) \right].
\label{eq:def-w-zeta}
\end{align}
We define the excess charge as 
\begin{align}
 \Delta q^\zeta(x)&=\frac{q^\zeta_0+q^\zeta_2(x)+2q^\zeta_w(x)}{2 |q^\zeta_{0}|}, \label{eq:Deltq-q-sig}
 \end{align}
In this case, we chose a different normalization as compared to the dielectric case because at zeroth order in lubrication, the local surface charge is not homogeneous and this would lead to unphysical contribution when assessing the global electroneutrality.  
Finally, using Eqs.~\eqref{eq:q0-zeta},\eqref{eq:q2-zeta},\eqref{eq:def-w-zeta}, Eq.~\eqref{eq:Deltq-q-sig} reduces to
\begin{align}
 \Delta  q^\zeta(x)& \simeq -\dfrac{\partial_x\left[\partial_x A_0^\zeta(x)\sinh(k_0 h(x))) \right]}{k_0^2\zeta \text{tanh}(k_0 h_0)} \,.
 \label{eq:Deltq-q-zeta}
\end{align}
As for the dielectric case, we note that Eq.~\eqref{eq:Deltq-q-zeta} shows the onset of local electroneutrality but also the fulfillment of global electroneutrality once $\Delta q^\zeta$ is integrated over the channel period. 
By comparing Fig.~\ref{fig:q_Eq} with Fig.~\ref{fig:q_Eq-z} we note that while for the dielectric channel there is a clear and sharp peak at the channel bottleneck, for the conducting channel the peaks are located where the slope of the channel walls is maximum, i.e., $x/L=\pm 0.25$. At the same time, on the top of the shift of the maxima, we also observe that the magnitude of the peak is reduced in the case of conducting as compared to dielectric channel walls.

\begin{figure*}[t]
 \includegraphics[scale=0.41]{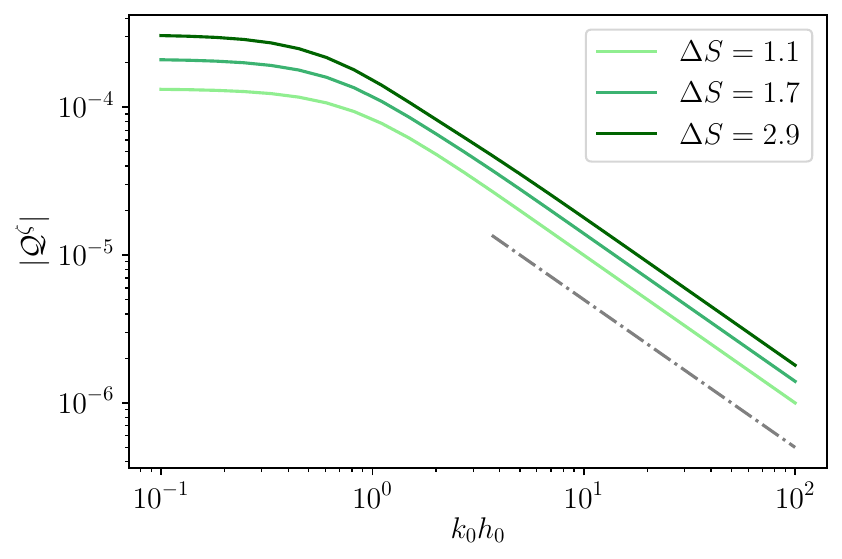}
 \includegraphics[scale=0.41]{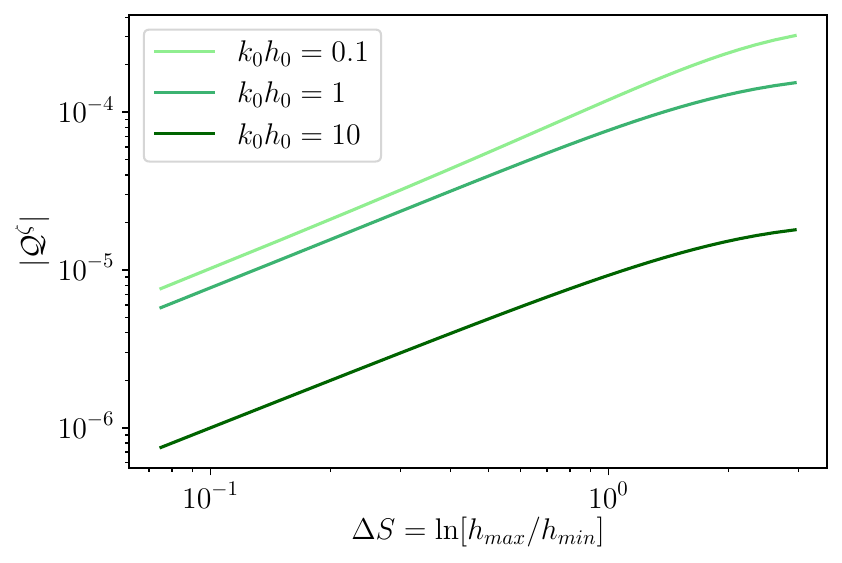}
 \includegraphics[scale=0.41]{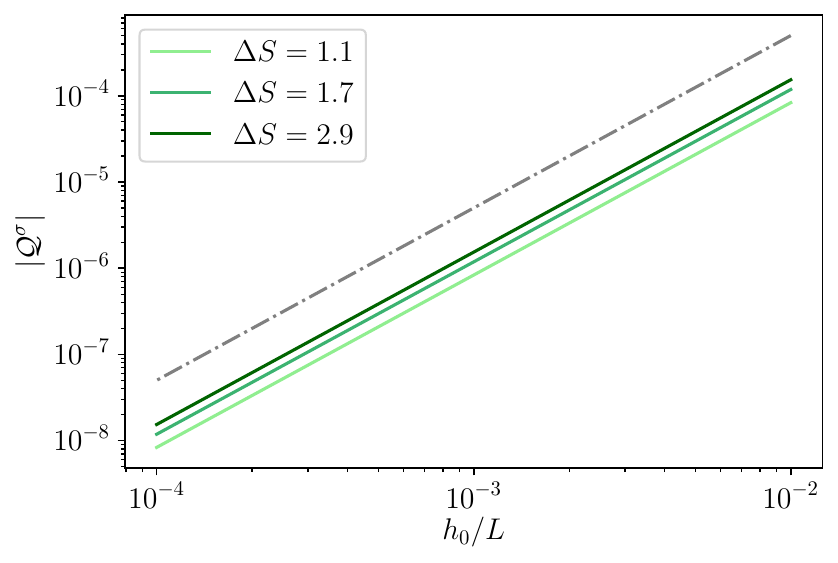}
  \includegraphics[scale=0.41]{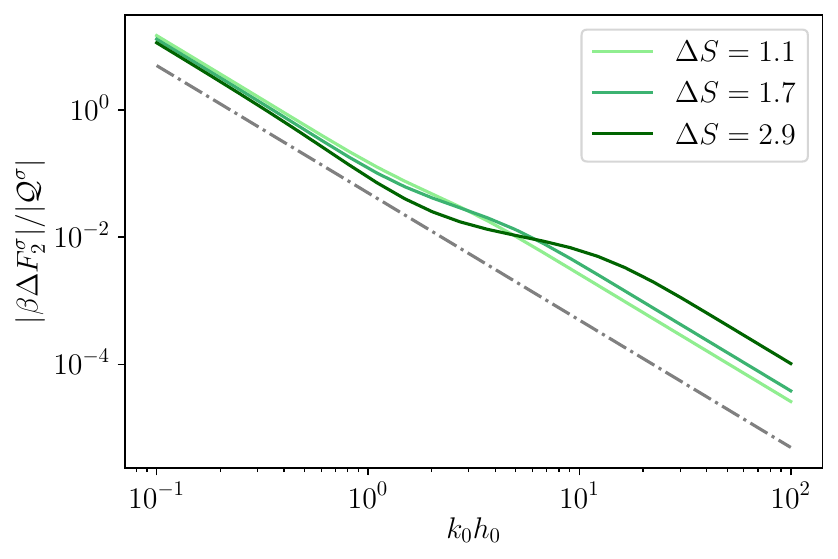}
 \includegraphics[scale=0.41]{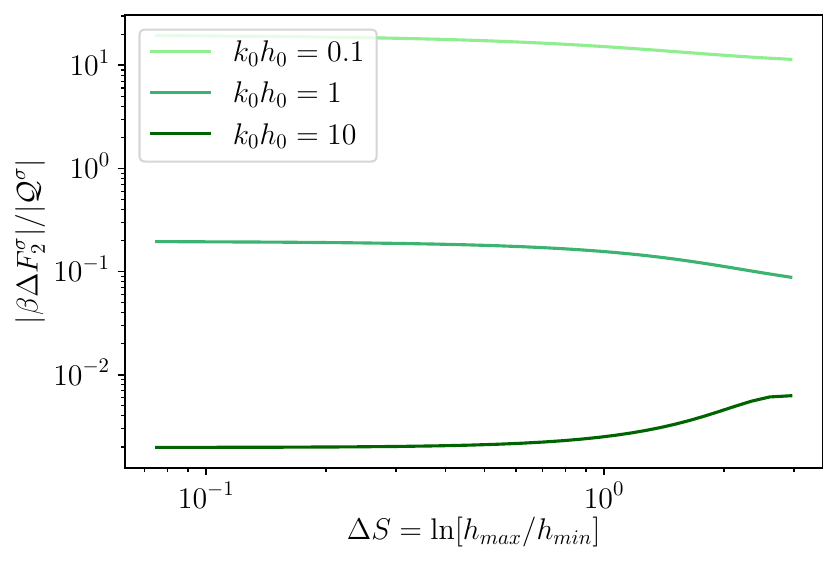}
 \includegraphics[scale=0.41]{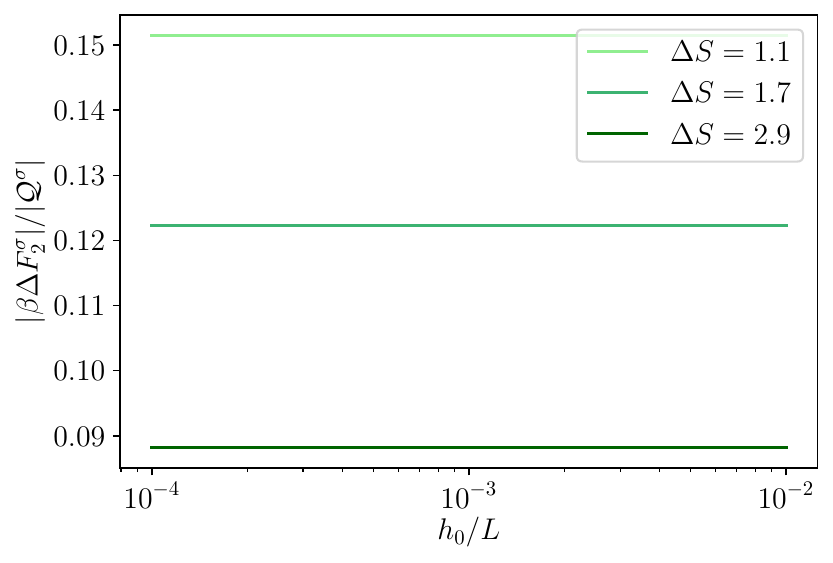}
 \caption{$2D$ Conducting channel. Top left: absolute value of the quadrupole moment as defined in Eq.~\eqref{eq:QQ}, $\mathcal{Q}^\sigma$ as a function of $k_0 h_0$ for different values of $\Delta S$ as reported in the legend and with $h_0/L=0.01$. The grey dashed line is proportional to $\propto (k_0 h_0)^{-1}$ and the dashed dotted to $\propto (k_0 h_0)^{-2}$. Top center: absolute value of the quadrupole moment, $\mathcal{Q}^\sigma$, as a function of $\Delta S$ for different magnitudes of $kh_0$ as reported in the legend and with $h_0/L=0.1$. The thin grey dotted line is a guide for the eye and it is proportional to $\Delta S^2$. Top right: absolute value of the quadrupole moment, $\mathcal{Q}^\sigma$, as a function of $h_0/L$ for different magnitudes of $\Delta S$ as reported in the legend and with $kh_0 = 1$. The thin grey dotted line is a guide for the eye and it is proportional to $(h_0/L)^2$. Bottom: ratio $\Delta F^\sigma_2/\mathcal{Q}^\sigma$ for the same values of the parameters of the corresponding top panel.
 }
 \label{fig:Q_Eq-z}
\end{figure*}

For what concerns the magnitude of the local excess charge captured by the quadrupolar moment, the top panel of Fig.~\ref{fig:Q_Eq-z} shows that $\mathcal{Q}^\zeta$ decays as $1/k_0 h_0$ for larger values of $k_0 h_0$ (as it is for dielectric channel walls) whereas for $k_0 h_0 \ll 1$, at variance with dielectric walls, $\mathcal{Q}^\zeta$ attains a plateau. Finally, the central and bottom panels of Fig.~\ref{fig:Q_Eq-z} show a behaviour similar to that observed for dielectric channel walls.   


\section{$2D$ free energy barrier}
While the leading term for the local electroneutrality breakdown is quadratic in $h_0/L$ this is not the case for the equilibrium free energy profile which, for a tracer ion with elementary charge $e$, reads
\begin{align}
\beta  F(x) = -\ln Z(x) = - \ln\left[\frac{1}{2h_0}\int\limits_{-h(x)}^{h(x)}e^{-\beta e \phi(x,y)}dy\right] 
\end{align}
and within the Debye-H\"uckel approximation it becomes
\begin{align}
\beta  F(x) \simeq \beta F_{DH}(x) =  -\ln\left[\frac{1}{2h_0}\int\limits_{-h(x)}^{h(x)}1-\beta e\phi(x,y)dy\right].
\label{eq:free-en}
\end{align}
By expanding the logarithm, Eq.~\eqref{eq:free-en} can be decomposed into the following contributions:
\begin{align}
\beta F_{DH}(x) \simeq \beta  F_{gas}(x)+\beta  F_{0}(x)+\beta  F_{2}(x), 
\end{align} 
with
\begin{align}
\beta  F_{gas}(x) &= -\ln\left[\frac{h(x)}{h_0}\right],\\
\beta  F_0(x) &= \frac{\beta e}{2h(x)}\int\limits_{-h(x)}^{h(x)}\phi_0(x,y)dy=-\frac{\beta e q_0(x)}{2h(x)\epsilon k_0^2},\\
\beta  F_2(x) &= \frac{\beta e}{2h(x)}\int\limits_{-h(x)}^{h(x)}\phi_2(x,y)dy=-\frac{\beta e q_2(x)}{2h(x)\epsilon k_0^2},
\end{align} 
where $\beta  F_{gas}(x)$ is the free energy profile of an uncharged point particle, whereas $\beta  F_0(x)$ and $\beta  F_2(x)$ are, respectively, the leading order and higher order correction for charged particles. The impact of the local excess charge on the dynamics of a tracer ion can be captured by the correction to the effective free energy barrier induced by the local excess charge
\begin{align}
\Delta F_2 = F_2(L/2)-F_2(0).
\end{align} 
In particular, we are interested in quantifying the contribution of the quadrupole moment to such a correction to the free energy barrier. The bottom rows of Fig.~\ref{fig:q_Eq-2},\ref{fig:Q_Eq-z} report the ratio between the second-order correction to the free energy difference and the quadrupole moment. 
As shown in the figures, the ratio of $\Delta F_2$ and $\mathcal{Q}$ is generally sensitive to both $k_0 h_0$ and $\Delta S$ hence highlighting the relevance of higher-order multipoles in the free energy difference. Finally, as expected $\Delta F_2$ and $\mathcal{Q}$ have the same scaling with $h_0/L$ and hence their ratio is insensitive to it.

\section{$3D$ first order lubrication approximation}
Exploiting the experience gathered for the $2D$ case we can straightforward write down the solution of the Debye-H\"uckel equation in the case of axially symmetric channels which, at first order in lubrication, reads
\begin{align}
\partial^2_r \phi_0(x,r)+\frac{1}{r}\partial_r\phi_0(x,r)=k_0^2 \phi_0(x,r)\,.
\end{align}
The solutions to it are
 \begin{align}
\phi^\zeta_0(x,r)&= \zeta\frac{I_0(k_0r)}{I_0(k_0 h(x))}, 
\label{eq:phi_zeta-3D} \\
\phi^\sigma_0(x,r)&= \frac{\sigma}{\epsilon k}\frac{I_0(k_0r)}{I_1(k_0 h(x))},
\label{eq:phi_sigm-3D}
\end{align}
where $I_n$ are modified Bessel functions of the first kind of order $n$. As for the $2D$ case, at linear order in the lubrication approximation the solution of the Debye-H\"uckel equation, Eq.~\eqref{eq:phi_sigm-3D}, fulfills local electroneutrality:
\begin{align}
q^\zeta_0(x) &= -2\pi \epsilon k_0^2 \int\limits_0^{h(x)} \phi^\zeta_0(x,r) r dr = -2\pi\epsilon \zeta\frac{k_0 h(x) I_1(k_0 h(x))}{I_0(k_0 h(x))},\label{eq:qw-z-0}\\
q^\sigma_0(x) &= -2\pi \epsilon k_0^2 \int\limits_0^{h(x)} \phi^\sigma_0(x,r) r dr = -2\pi h(x)\sigma\,.
\end{align}
This indeed is minus the charge on the wall.
\section{$3D$ higher order corrections}
At second order in lubrication the Debye-H\"uckel equation becomes
\begin{align}
\partial^2_x \phi_0(x,r)+\partial^2_r \phi_2(x,r)+\frac{1}{r}\partial_r\phi_2(x,r)=k_0^2 \phi_2(x,r)\,,
\end{align}
whose solution reads\footnote{We recall the following property of the modified Bessel function of the first kind $\partial_z(z^n I_n(z))=z^n I_{n-1}(z)$}
\begin{align}
\phi_2(x,r) = B_2(x)I_0(k_0 r)-\frac{\partial_x^2 B_0 (x)}{2k_0^2}  \left(I_0(k_0r)+k_0rI_1(k_0 r)\right),
\label{eq:sol-2nd}
\end{align}
with
\begin{align}
B^\zeta_0(x) &= \frac{\sigma}{\epsilon k_0}\frac{1}{I_0(k_0 h(x))},\\
B^\sigma_0(x) &= \frac{\sigma}{\epsilon k_0}\frac{1}{I_1(k_0 h(x))}
\end{align}
and $B_2$ is determined by the boundary conditions. 
\section{$3D$ local electroneutrality breakdown}
\paragraph{Dielectric channel walls}
For dielectric channel walls the boundary condition is the same as the one derived for the $2D$ case, Eq.~\eqref{eq:approx-BC}, and reads
\begin{align}
\partial_r \phi^\sigma_2(x,r)&= \frac{1}{2}\frac{\sigma}{\epsilon}(\partial_x h(x))^2\nonumber\\
&-\frac{\sigma}{\epsilon k_0} \frac{I_0(k_0 h(x))}{I^2_1(k_0 h(x))}\partial_xI_1(k_0 h(x))\partial_x h(x).
\end{align}
Hence, 
\begin{align}
B^\sigma_2(x) =& \frac{1}{2}B^\sigma_0(x)(\partial_x h(x))^2+\frac{I_0(k_0 h(x))}{k_0 I_1(k_0 h(x))}\partial_x B_0\partial_x h(x)\nonumber\\
&+\frac{\partial_x^2 B_0 (x)}{2k_0^2}  \left[1+k_0h(x)\frac{I_0(k_0h(x))}{I_1(k_0h(x))}\right].
\label{eq:B2}
\end{align}
Accordingly, the second-order correction to the integrated local charge can be obtained by substituting Eq.~\eqref{eq:B2} into Eq.~\eqref{eq:phi_sigm-3D} and integrating along the radial direction (see appendix B):
\begin{align}
q^\sigma_2(x) &= -2\pi\epsilon k_0^2 \int_0^{h(x)} \phi_2(x,r)r dr\nonumber\\ 
&= -\pi\sigma (\partial_x h(x))^2 -2\pi \epsilon I_0(k_0 h(x))h(x)\partial_x h(x)\partial_x B_0(x) \\
&-2\pi\epsilon \frac{\partial_x^2 B_0 (x)}{2k_0^2} (k_0 h(x))^2 \left(I_0(k_0 h(x))-I_2(k_0 h(x))\right)\nonumber
\label{eq:dq-3D-0}
\end{align}
Finally, recalling that, at second order, the surface charge per unit longitudinal length is
\begin{align}
q^\sigma_w = 2\pi \sigma \left(1+\frac{1}{2}(\partial_x h(x))^2\right),
\end{align}
the local charge excess, at second order in lubrication, reads (see appendix A) 
\begin{align}
\Delta & q^\sigma(x)= -\partial_x \left[\frac{\partial_x B_0 (x)}{\sigma k_0/\epsilon}h(x) I_1(k_0 h(x))\right].
\label{eq:dq-3DDD}
\end{align}
\begin{figure}[t]
 \includegraphics[scale=0.5]{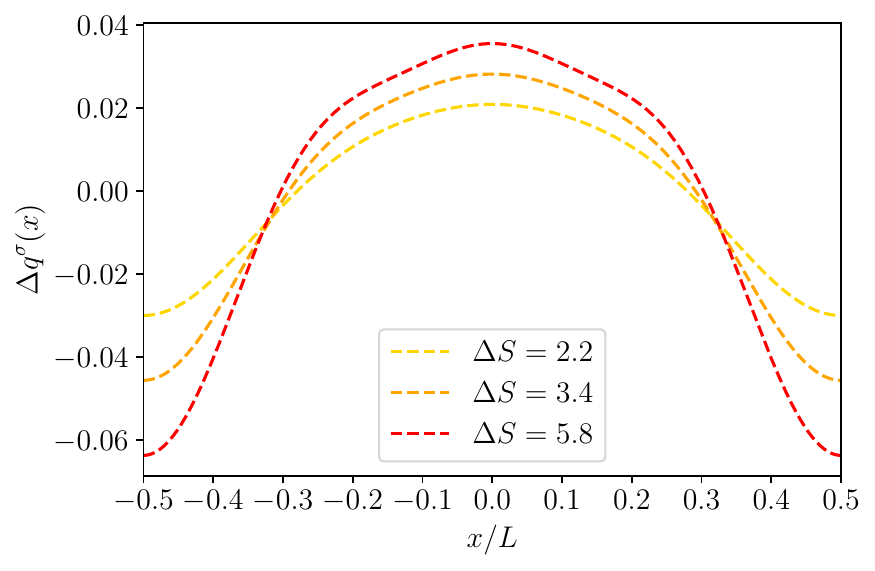}
 \caption{$3D$ Dielectric channel. Local excess charge $\Delta q(x)$  as a function of the normalized position $x/L$ along the channel axis for different values of $\Delta S$ as deailed in the legend with $h_0/L=0.1$ and $kh_0=1$. }
 \label{fig:q_Eq-3D}
\end{figure}
\begin{figure*}[t]
 \includegraphics[scale=0.41]{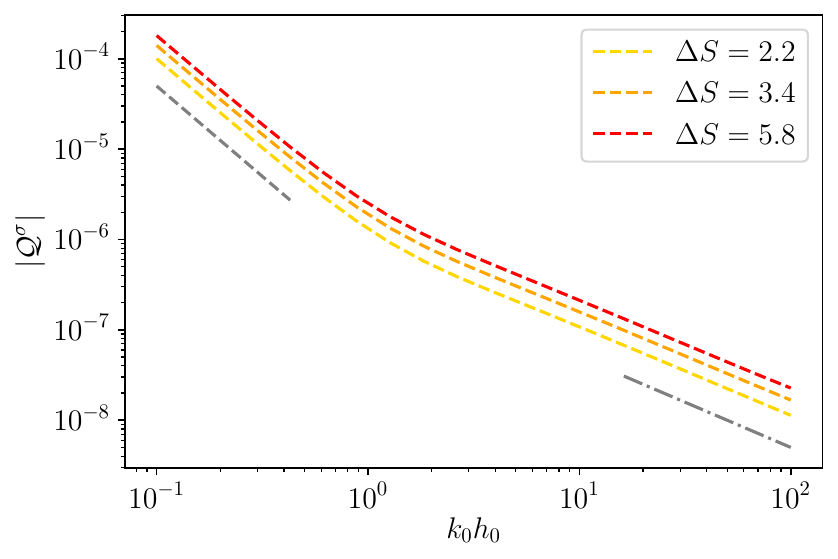}
 \includegraphics[scale=0.41]{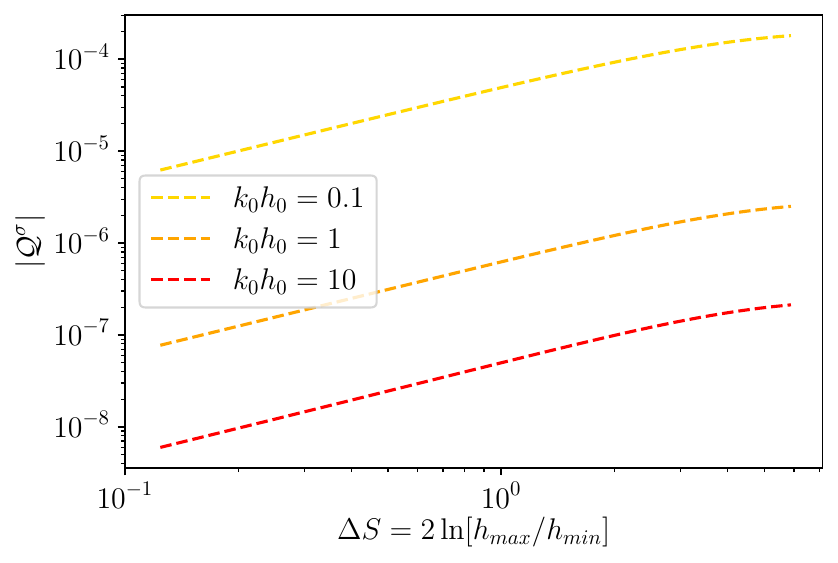}
 \includegraphics[scale=0.41]{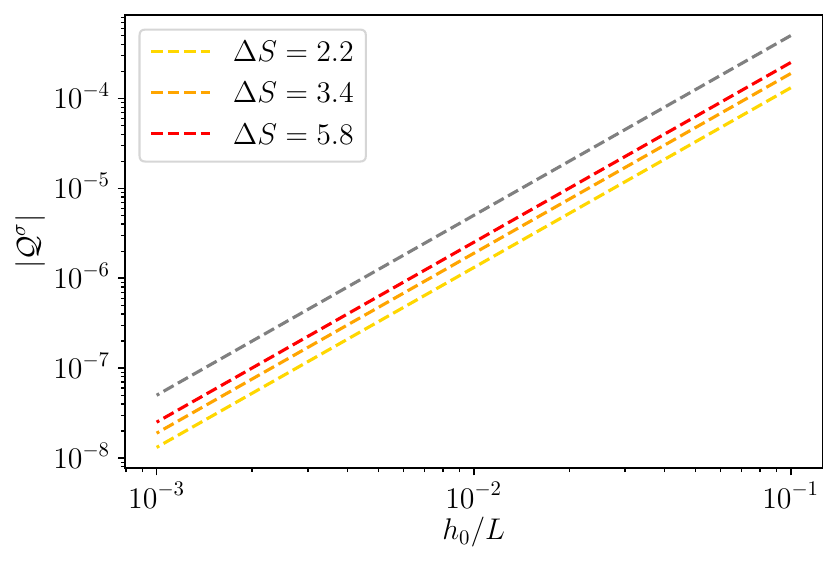}
 \includegraphics[scale=0.41]{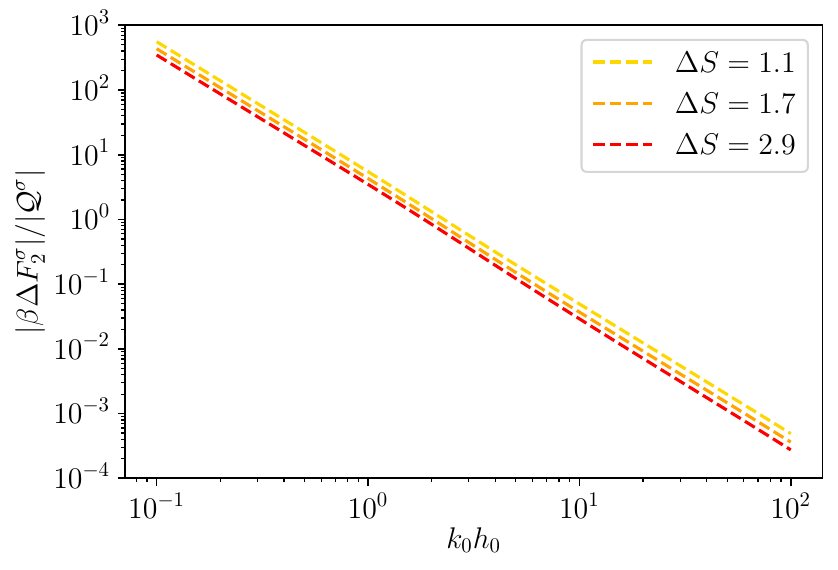}
 \includegraphics[scale=0.41]{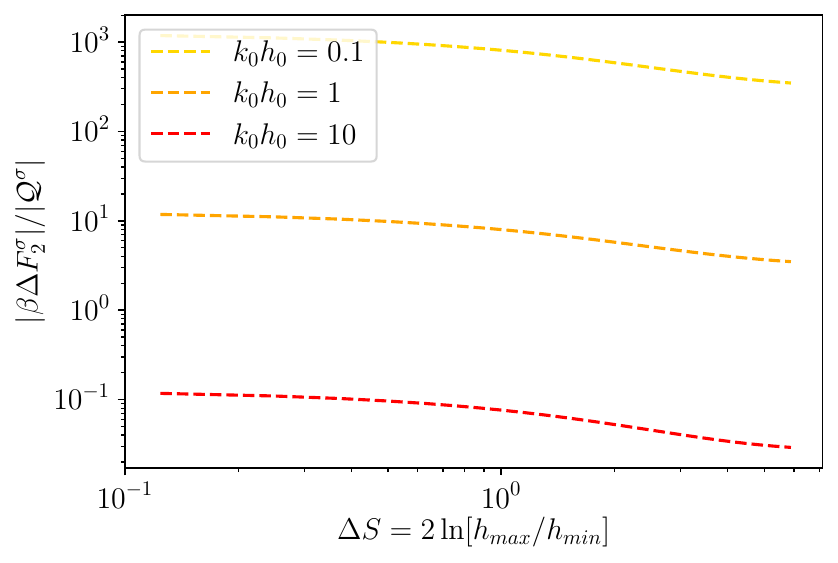}
 \includegraphics[scale=0.41]{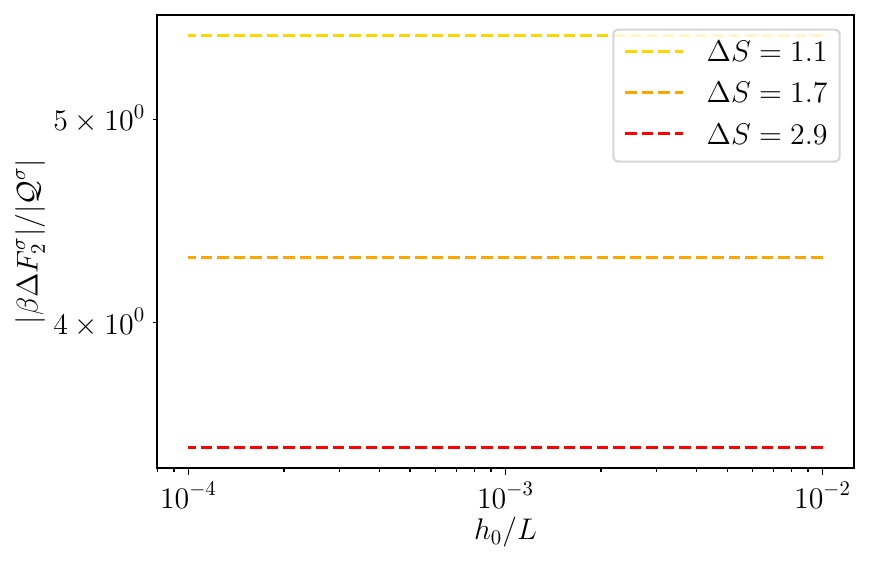}
 \caption{$3D$ Dielectric channel. Top left: local excess charge  $\Delta q^\sigma$ as a function of $k_0 h_0$ for different values of $\Delta S$ as reported in the legend and with $h_0/L=0.01$. The grey dashed line is proportional to $\propto (k_0 h_0)^{-1}$ and the dashed dotted to $\propto (k_0 h_0)^{-2}$. Top center: local excess charge $\Delta q^\sigma$ as a function of $\Delta S$ for different magnitudes of $kh_0$ as reported in the legend and with $h_0/L=0.1$. Top right: local excess charge $\Delta q^\sigma$ as a function of $h_0/L$ for different magnitudes of $\Delta S$ as reported in the legend and with $kh_0 = 1$. The thin grey dotted line is a guide for the eye and it is proportional to $(h_0/L)^2$. Bottom: ratio $\Delta F^\sigma_2/\mathcal{Q}^\sigma$ for the same values of the parameters of the corresponding top panel.}
 \label{fig:Q_Eq-3D}
\end{figure*}

Fig.~\ref{fig:q_Eq-3D} shows the dependence of $\Delta q$ on the position. Interestingly, as for the $2D$ case, $\Delta q$ displays a maximum at the channel bottleneck, $x/L=0.5$. However, a part of the different shape of the profile, the main striking difference between Fig.~\ref{fig:q_Eq-3D} and Fig.~\ref{fig:q_Eq} is the difference in magnitude of the effect. In fact, while for the $2D$ case the excess charge at the peak is comparable to the net charge density on the wall, for the $3D$ case (with similar geometry, $\Delta S$, and Debye length, $k_0 h_0$) the effect is weaker. Hence we do expect that local electroneutrality breakdown to be more prominent for slab-like channels then for cylindrical pores. More in detail, Fig.~\ref{fig:Q_Eq-3D} shows again a similar trend as compared to the $2D$ case, Fig.~\ref{fig:q_Eq-2}, the only major difference being the non-monotonous dependence on $\Delta S$  shown for large values of $k_0 h_0$.
\begin{figure}[t]
 \includegraphics[scale=0.5]{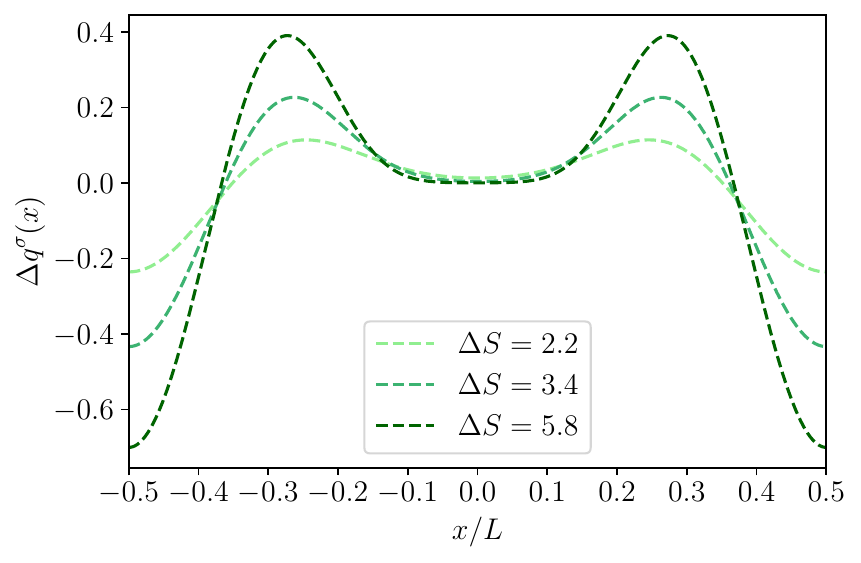}
 \caption{$3D$ Conducting channel. Local excess charge $\Delta q(x)$  as a function of the normalized position $x/L$ along the channel axis for different values of $\Delta S$ as detailed in the legend with $h_0/L=0.1$ and $kh_0=1$. }
 \label{fig:q_Eq-3D-zeta}
\end{figure}
\begin{figure*}[t]
 \includegraphics[scale=0.41]{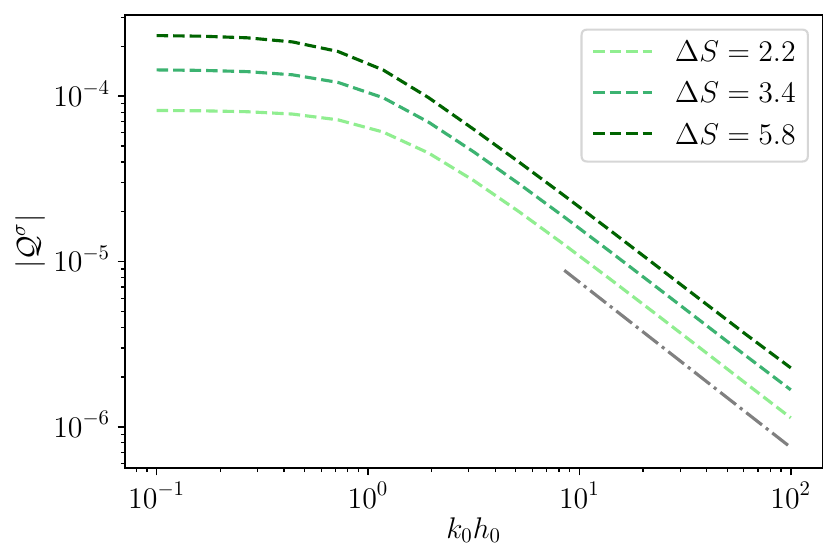}
 \includegraphics[scale=0.41]{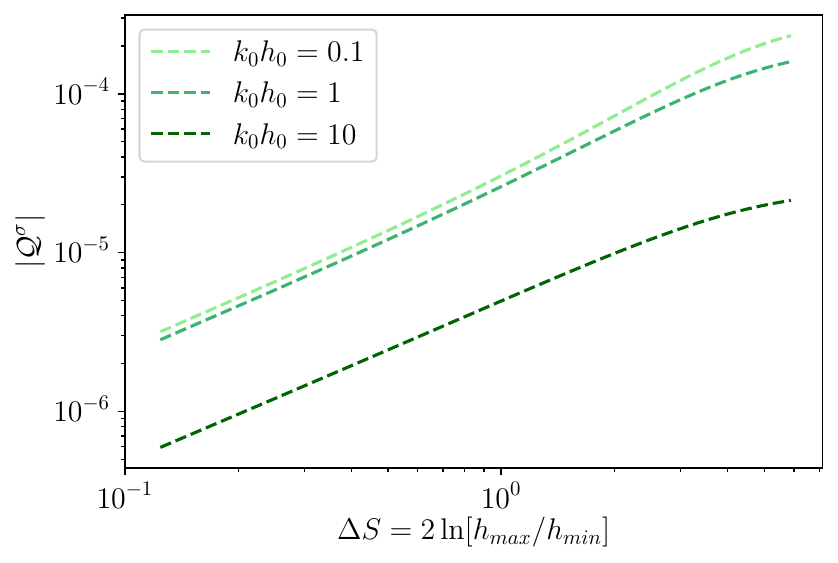}
 \includegraphics[scale=0.41]{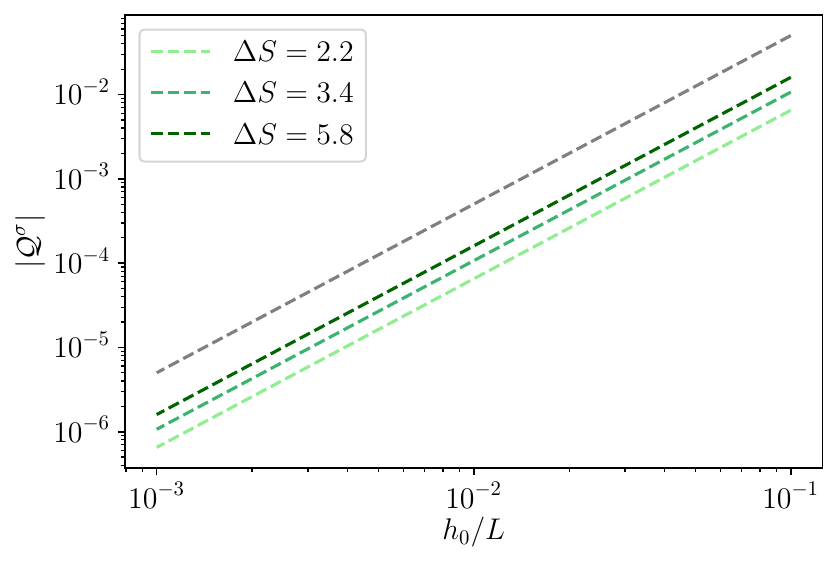}
 \includegraphics[scale=0.41]{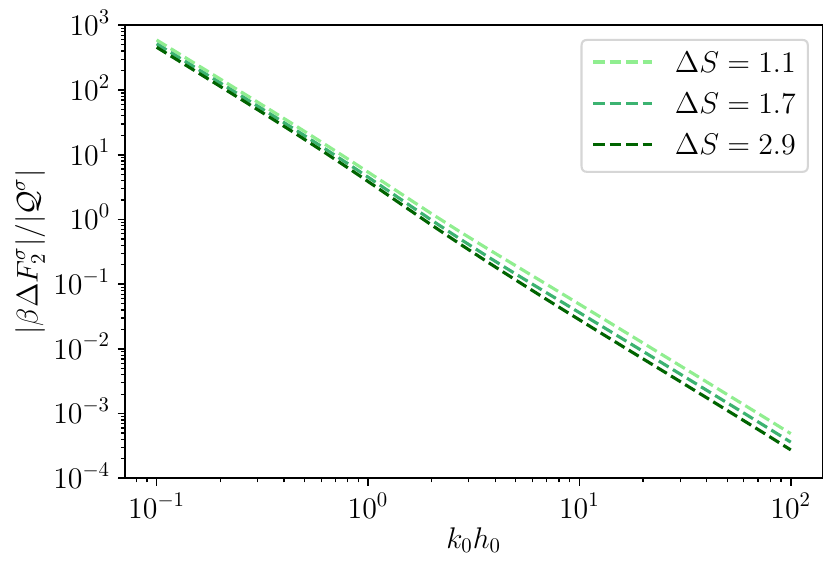}
 \includegraphics[scale=0.41]{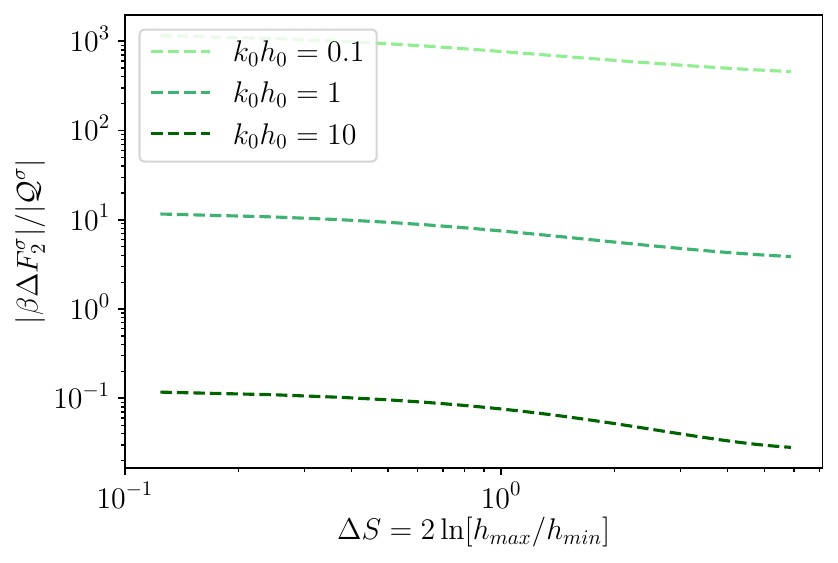}
 \includegraphics[scale=0.41]{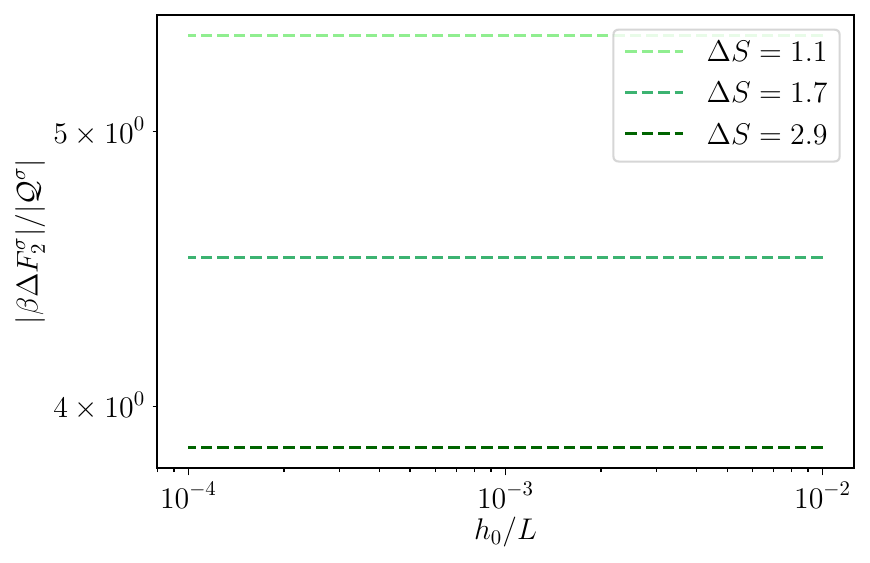}
 \caption{$3D$ Conducting channel. Top: local excess charge  $\Delta q^\sigma$ as a function of $k_0 h_0$ for different values of $\Delta S$ as reported in the legend and with $h_0/L=0.01$. The grey dashed line is proportional to $\propto (k_0 h_0)^{-1}$ and the dashed dotted to $\propto (k_0 h_0)^{-2}$.  Center: local excess charge $\Delta q^\sigma$ as a function of $\Delta S$ for different magnitudes of $kh_0$ as reported in the legend and with $h_0/L=0.1$. Bottom: local excess charge $\Delta q^\sigma$ as a function of $h_0/L$ for different magnitudes of $\Delta S$ as reported in the legend and with $kh_0 = 1$. The thin grey dotted line is a guide for the eye and it is proportional to $(h_0/L)^2$. Bottom: ratio $\Delta F^\sigma_2/\mathcal{Q}^\zeta$ for the same values of the parameters of the corresponding top panel.}
 \label{fig:Q_Eq-3D-z}
\end{figure*}

\paragraph{Conducting channel walls}
For conducting channel walls the boundary condition is the same as the one derived for the $2D$ case, namely $\phi_2^\zeta(x,h(x))=0$, and hence Eq.~\eqref{eq:sol-2nd} leads to
\begin{align}
B^\zeta_2(x) =& \frac{\partial_x^2 B_0 (x)}{2k_0^2}  \left[1+k_0h(x)\frac{I_1(k_0h(x))}{I_0(k_0h(x))}\right]\,.
\label{eq:B2-z}
\end{align}
The second order correction to the wall charge can be obtained as in the $2D$ case, Eq.~\eqref{eq:qw_z-2}, where we change $y$ for $r$ and we multiply by the local area $2\pi h(x)(1+1/2(\partial_x h(x))^2)$:
\begin{align}\label{eq:qw-z}
q^\zeta_w =& -2\pi h(x)\epsilon \left[I_0(k_0 h(x))\partial_xB_0^\zeta(x)\partial_x h(x)\right.\nonumber\\
&-k_0\zeta\frac{I_1(k_0 h(x))}{I_0(k_0 h(x))}-B_2(x)k_0I_1(k_0 h(x))\\
&\left.+\frac{\partial_x^2B_0^\zeta(x)}{2k_0}\left(k_0 h(x)I_0(k_0 h(x))+I_1(k_0 h(x))\right) \right]  \nonumber
\end{align}
Similarly to the $2D$ case, the net charge in the liquid phase reads
\begin{align}\label{eq:q2-z}
q_2^\zeta &= -2\pi\epsilon k_0^2 \int_0^{h(x)}\phi_2(x,r)r dr\nonumber\\
=& -2\pi\epsilon k_0 h(x) \Big[B_2(x)I_1(k_0 h(x))\\
&-\frac{1}{2k_0^2}\partial_x^2B_0(x)\left(I_1(k_0 h(x))+k_0 h(x) I_2(k_0 h(x))\right)\Big].\nonumber
\end{align}  
Combining Eq.~\eqref{eq:qw-z-0}, Eq.~\eqref{eq:qw-z} and Eq.~\eqref{eq:q2-z} leads to the local excess charge density
\begin{align}
\Delta q^\zeta =-\frac{2}{k^2_0} \dfrac{\partial_x\left[k_0h(x)I_1(k_0 h(x))\partial_x B_0^\zeta(x)\right]}{\zeta k_0 h_0 \frac{I_1(k_0h_0)}{I_0(k_0h_0)}}.
\end{align}

\section{$3D$ free energy barrier}
Similarly to the $2D$ case, at equilibrium, the local free energy of a tracer ion with elementary charge $e$ is given by 
\begin{align}
\beta  F(x) = -\ln Z(x) = - \ln\left[\frac{1}{\pi h^2_0}\int\limits_{0}^{h(x)}e^{-\beta e \phi(x,r)}rdr\right] \,.
\label{eq:free-en-3D}
\end{align}
Within the Debye-H\"uckel approximation and expanding the logarithm, Eq.~\eqref{eq:free-en} can be decomposed into the following contributions:
\begin{align}
\beta F_{DH}(x) \simeq \beta  F_{gas}(x)+\beta  F_{0}(x)+\beta  F_{2}(x) ,
\end{align} 
with
\begin{align}
\beta  F_{gas}(x) &= -2\ln\left[\frac{h(x)}{h_0}\right],\\
\beta  F_0(x) &= \frac{\beta e}{\pi h^2(x)}\int\limits_{-h(x)}^{h(x)}\phi_0(x,y)dy=-\frac{\beta e q_0(x)}{\pi \epsilon  k_0^2 h^2(x)},\\
\beta  F_2(x) &= \frac{\beta e}{\pi h^2(x)}\int\limits_{-h(x)}^{h(x)}\phi_2(x,y)dy=-\frac{\beta e q_2(x)}{\pi \epsilon  k_0^2 h^2(x)},
\end{align} 
where $\beta  F_{gas}(x)$ is the free energy profile of an uncharged point particle whereas $\beta  F_0(x)$ and $\beta  F_2(x)$ are, respectively, the leading order and higher order correction for charged particles. In order to assess the impact of the local excess charge on the dynamics of a tracer ion, the bottom rows of Fig.~\ref{fig:q_Eq-2},\ref{fig:Q_Eq-z}  report the ratio between the second-order correction to the free energy difference 
\begin{align}
\Delta F_2 = F_2(L/2)-F_2(0)
\end{align} 
and the quadrupole moment. As already mentioned for the quadrupole, the overall behaviour of $\Delta F_2$ resembles the one observed in the respective $2D$ cases.

\section{Conclusions}
In this contribution, we focus on the case of an electrolyte embedded between corrugated channel walls. In order to gain analytical insight we restrict our analysis to channels whose section is varying smoothly enough so that we can exploit the lubrication approximation to solve for the Poisson equation. 
Under such approximation, we have derived closed formulas for the corrections induced by the varying section of the channel to the local charge distribution. In particular, at equilibrium, the Debye length keeps homogeneous even when second-order corrections in the lubrication expansion are accounted for. At variance, while at first order in lubrication the local electroneutrality of the system is recovered, this is not so for second order corrections. This implies that upon reducing the length scale separation between the longitudinal and transverse direction the local excess charge will grow and this will also induce additional corrections to the effective free energy profile experienced by a tracer ion. Such local charge reorganization within corrugated channels has been observed so far only in out of equilibrium situations~\cite{Chinappi2018} where the advection of the ions plays a major role. Our results show that such a phenomenon occurs also at equilibrium and hence solely due to the interplay between the geometry of the channel and the electrostatic forces. 
In order to assess the robustness of our results we have derived such corrections for both dielectric and conducting channel walls in both planar ($2D$) and cylindrical ($3D$) geometries. Indeed, our results show quite a remarkable similarity between the $2D$ and $3D$ cases for both dielectric and conducting channel walls. Interestingly, the dielectric case shows an enhanced sensitivity to the dimensionality, as compared to the conducting case. In particular, the local (integrated) excess charge attains its maximum at the channel bottleneck ($x/L=0$), for $2D$ dielectric walls, and it can be as large as the bare charge (density) on the walls. This is not the case for cylindrical channels ($3D$) for which the local excess charge is $\simeq 100$ times smaller than the local charge. 
The difference in the location of the excess charge along the channel for dielectric and conducting walls indicates that the specific boundary conditions play a relevant role in the transport properties of confined electrolytes and ions.
At variance, for conducting channels, the maxima of the local excess charge are located where the slope of the channel walls is maximum ($x/L=\pm 0.25$) for both $2D$ and $3D$ cases. 
All in all, the magnitude of the corrections that we report on are not very large and indeed this is expected since they are obtained via an expansion. However, this may not be the case when the longitudinal length is comparable to the Debye length. For typical values of the Debye length, $\lambda \simeq 10-100$nm, this implies to have channels with length $L\lesssim 100$nm. This is the case for many biological ionic channels as well as synthetic pores and membranes.

\section{acknowledgments}
P.M and J.H. acknowledge funding by the Deutsche Forschungsgemeinschaft (DFG, German Research Foundation) Project-ID 416229255—SFB 1411.
I.P. acknowledges support from Ministerio de Ciencia, Innovaci\'on y Universidades MCIU/AEI/FEDER for financial support under 
grant agreement PID2021-126570NB-100 AEI/FEDER-EU, from Generalitat de Catalunya  under Program Icrea Acad\`emia and project 2021SGR-673. 

\pagebreak

\bibliography{bib_paper_entropy,from_zotero}

\appendix

\begin{widetext}

\section{Derivation of Eq.~\eqref{eq:dq-3D-0}}

In order to derive Eq.~\eqref{eq:dq-3D-0} we exploit the following relations
\begin{align}
\int_0^{h(x)} I_0 (k_0 r) r dr &= \frac{1}{k_0^2}\int_0^{k_0 h(x)} I_0 (z) z dz =   \frac{1}{k_0^2}k_0 h(x) I_1 (k_0 h(x))\\
\int_0^{h(x)} k_0 r I_1(k_0 r) r dr &= \frac{1}{k_0^2}\int_0^{k_0 h(x)} I_1 (z) z^2 dz = \frac{1}{k_0^2}(k_0 h(x))^2 I_2 (k_0 h(x))
\end{align}
were we used $z=k_0 r$. 
Finally we get:
\begin{align}
q_2(x) &= -2\pi\epsilon k_0^2 \int_0^{h(x)}\phi_2(x,r)r dr\nonumber\\
&= -2\pi\epsilon k_0 h(x) \left[B_2(x)I_1(k_0 h(x))-\frac{1}{2k_0^2}\partial_x^2B_0(x)\left(I_1(k_0 h(x))+k_0 h(x) I_2(k_0 h(x))\right)\right]
\end{align}
By substituting the relative expressions for $B_2(x)$ and $B_0(x)$ we get
\begin{align}
q_2^\sigma(x) &= -\pi\sigma h(x) -2\pi \epsilon I_0(k_0 h(x))h(x)\partial_x h(x)\partial_x B_0(x) -2\pi\epsilon \frac{\partial_x^2 B_0 (x)}{2k_0^2} (k_0 h(x))^2 \left(I_0(k_0 h(x))-I_2(k_0 h(x))\right)\\
q_2^\zeta(x) &= -2\pi \epsilon \frac{\partial_x^2 B_0(x)}{2k_0^2}\left(k_0 h(x)\right)^2\left[\frac{I_1^2(k_0 h(x))}{I_0(k_0 h(x))}-I_2(k_0 h(x))\right]
\end{align}
Finally, using the recursive relation 
\begin{align}
\partial_x I_0(k_0 h(x))= I_1(k_0 h(x))k_0 \partial_x h(x)
\end{align}
we obtain
\begin{align}
q_2^\sigma(x) &= -\pi\sigma h(x) -2\pi \epsilon I_0(k_0 h(x))h(x)\partial_x h(x)\partial_x B_0(x) +2\pi\epsilon \frac{\partial_x^2 B_0 (x)}{k_0^2} k_0 h(x) I_1(k_0 h(x))
\end{align}

\section{Derivation of Eq.~$(64)$}

We want to show that 
\begin{align}
\partial_x &\left[\frac{\partial_x B_0 (x)}{2k_0^2}(k_0 h(x))^2  \left(I_0(k_0 h(x))-I_2(k_0 h(x))\right)\right] =\nonumber\\
& =- I_0(k_0 h(x))h(x)\partial_x h(x)\partial_x B_0(x) - \frac{\partial_x^2 B_0 (x)}{2k_0^2} (k_0 h(x))^2 \left(I_1(k_0 h(x))-I_2(k_0 h(x))\right)
\end{align}
that reduces to showing the following relation
\begin{align}
\partial_x &\left[(k_0 h(x))^2  \left(I_0(k_0 h(x))-I_2(k_0 h(x))\right)\right] =- 2I_0(k_0 h(x))h(x)\partial_x h(x)\label{app:deriv_B0-0}
\end{align}
In order to do so we use the following relation
\begin{align}
\partial_x (x^n I_n(x))= x^n I_{n-1}(x)
\end{align}
that leads to 
\begin{align}
\partial_x\left[(k_0 h(x))^2 I_2(k_0 h(x))\right] = (k_0 h(x))^2 I_1(k_0 h(x))\partial_x(k_0 h(x))\label{app:deriv_B0-1}
\end{align}
and 
\begin{align}
\partial_x\left[(k_0 h(x))^2 I_0(k_0 h(x))\right] &=   2 k_0 h(x) I_0(k_0 h(x)) \partial_x (k_0 h(x)) + (k_0 h(x))^2 \partial_x I_0(k_0 h(x)))\nonumber\\
&= 2k_0 h(x) I_0(k_0 h(x))\partial_x (k_0 h(x))+ (k_0 h(x))^2 I_1(k_0 h(x))\partial_x (k_0 h(x))\label{app:deriv_B0-2}
\end{align}
where in the last step we used 
\begin{align}
I_{n+1}(x) = I_{n-1} - \frac{2n}{x}I_n(x)
\end{align}
Summing Eqs.~\eqref{app:deriv_B0-1},\eqref{app:deriv_B0-2} we get Eq.~\eqref{app:deriv_B0-0} and hence Eq.~($64$).

\end{widetext}

\end{document}